\documentclass[twocolumn]{autart}    

\usepackage{graphicx}          

\usepackage{amsmath}
\usepackage{mathtools}
\usepackage{amsfonts}
\usepackage{amssymb}
\usepackage{algorithm}
\usepackage{algpseudocode}
\usepackage{subfig}
\usepackage{soul}
\usepackage{color}
\usepackage{natbib} 

\begin{document}

\begin{frontmatter}

\title{Provably Correct Training of Neural Network\\ Controllers Using Reachability Analysis}                                       





\thanks[footnoteinfo]{This  work  was  partially  sponsored  by  the  NSF  awards \#CNS-2002405 and \#CNS-2013824. \\
This paper was not presented at any conference.
}

\author[uci]{Xiaowu Sun}\ead{xiaowus@uci.edu},     
\author[uci]{Yasser Shoukry}\ead{yshoukry@uci.edu},               

\address[uci]{Department of Electrical Engineering and Computer Science, University of California, Irvine, CA 92697 USA}                                          


\begin{abstract}                          
In this paper, we consider the problem of training neural network (NN) controllers for nonlinear dynamical systems that are guaranteed to satisfy safety and liveness (e.g., reach-avoid) properties. Our approach is to combine model-based design methodologies for dynamical systems with data-driven approaches to achieve this target. We confine our attention to NNs with Rectifier Linear Unit (ReLU) nonlinearity which are known to represent Continuous Piece-Wise Affine (CPWA) functions. Given a mathematical model of the dynamical system, we compute a finite-state abstract model that captures the closed-loop behavior under all possible CPWA controllers. Using this finite-state abstract model, our framework identifies a family of CPWA functions guaranteed to satisfy the safety requirements. We augment the learning algorithm with a NN weight projection operator during training that enforces the resulting NN to represent a CPWA function from the provably safe family of CPWA functions. Moreover, the proposed framework uses the finite-state abstract model to identify candidate CPWA functions that may satisfy the liveness properties. Using such candidate CPWA functions, the proposed framework biases the NN training to achieve the liveness specification. We show the efficacy of the proposed framework both in simulation and on an actual robotic vehicle.
\end{abstract}

\end{frontmatter}

\section{Introduction} 
The last decade has witnessed tremendous success in using machine learning (ML) in a multitude of safety-critical cyber-physical systems domains, such as autonomous vehicles, drones, and smart cities. Indeed, end-to-end learning is attractive for the realization of feedback controllers for such complex cyber-physical systems, thanks to the appeal of designing control systems based on purely data-driven architectures. However, regardless of the explosion in the use of machine learning to design data-driven feedback controllers, providing formal safety and reliability guarantees of these ML-based controllers is in question. It is then unsurprising the recent focus in the literature on the problem of safe and trustworthy autonomy in general, and safe reinforcement learning, in particular.

The literature on the safe design of ML-based controllers for dynamical and hybrid systems can be classified according to three broad approaches, namely (i) incorporating safety in the training of ML-based controllers, (ii) post-training verification of ML-based controllers, and (iii) online validation of safety and control intervention. 
Representative examples of the first approach include     
reward-shaping~\citep{stone2020reward,saunders2018trial}, Bayesian and robust   
regression~\citep{berkenkamp2016bayesian,liu2019robust,pauli2020training}, and policy optimization with 
constraints~\citep{abbeel2017icml,turchetta2016safe,wen2020safe}.
Unfortunately, these approaches do not provide provable guarantees on the safety of the trained controller. 

To provide strong safety and reliability guarantees, several works in the literature focus on applying formal verification techniques (e.g., model checking) to verify pre-trained ML-based controllers' formal safety properties. Representative examples of this approach are the use of SMT-like  solvers~\citep{dutta2018output,liu2019algorithms,sun2019formal} and 
hybrid-system verification~\citep{fazlyab2019efficient,ivanov2019verisig,xiang2019reachable}. However, these techniques only assess a given ML-based controller's safety rather than design or train a safe agent.

Due to the lack of safety guarantees on the resulting ML-based controllers, researchers proposed several techniques to \emph{restrict} the output of the ML-based controller to a set of safe control actions. 
Such a set of safe actions can be obtained through Hamilton-Jacobi analysis~\citep{fisac2018general,fisac2021rss}, and barrier certificates~\citep{abate2021hscc,pappas2021hscc,cheng2019end,ferlez2020shieldnn,li2019temporal,matni2020cdc,taylor2020control,wabersich2018scalable,wang2018safe,cassandras2020adaptivecbf}. 
Unfortunately, methods of this type suffer from being computationally expensive, specific to certain controller structures or else employ training algorithms that require certain assumptions on the system model. 
Other techniques in this domain include synthesizing a safety layer or shield based on safe model predictive control with the assumption of existing a terminal safe set~\citep{bastani2021rss,zeilinger2018cdc,zeilinger2021filter}, and Lyapunov methods~\citep{berkenkamp2017safe,chow2018lyapunov,chow2019lyapunov} which focus mainly on providing stability guarantees rather than general safety and liveness guarantees.

This paper proposes a principled framework combining model-based control and data-driven neural network training to achieve enhanced reliability and verifiability. Our framework bridges ideas from reachability analysis to guide and bias the neural network controller's training and provides strong reliability guarantees. Our starting point is the fact that Neural Networks (NNs) with Rectifier Linear Unit (ReLU) nonlinearity represent Continuous Piece-Wise Affine (CPWA) functions. Given a nonlinear model of the system, we compute a finite-state abstract model capable of capturing the closed-loop behavior under \emph{all} possible CPWA controllers. Such a finite-state abstract model can be computed using a direct extension of existing reachability tools. Next, our framework uses this abstract model to search for a family of \emph{safe} CPWA controllers such that the system controlled by any controller from this family is guaranteed to satisfy the safety properties. During the neural network training, we use a novel \emph{projection} operator that projects the trained neural network weights to ensure that the resulting NN represents a CPWA function from the identified family of safe CPWA functions.

Unlike the safety property, satisfying the liveness property can not be enforced by projecting the trained neural network weights. Therefore, to account for the liveness properties, our framework utilizes the abstract model further to refine the family of safe CPWA functions to find \emph{candidate} CPWA functions that may satisfy the liveness properties. The framework then ranks these \emph{candidates} and biases the NN training accordingly until a NN that satisfies the liveness property is obtained.
%
%
In conclusion, the contributions of this paper can be summarized as follows:
\begin{enumerate}
    \item An abstraction-based framework that captures the behavior of \emph{all} neural network controllers.
    \item A novel projected training algorithm to train provably safe NN controllers.
    \item A procedure to bias the NN training to satisfy the liveness properties. 
\end{enumerate}


\section{Problem Formulation} 
\textbf{Notation:}
The symbols $\mathbb{R}$ and $\mathbb{N}$ denote the set of real and natural numbers, respectively. The symbols $\land$ and $\Longrightarrow$ denote the logical AND and logical IMPLIES, respectively. We denote by $\mathrm{Int}(\mathcal{S})$ the interior of a set $\mathcal{S}$ and by $\mathrm{Vert}(\mathcal{R})$ the the set of all vertices of a polytope $\mathcal{R}$. 
We use $\Psi_\text{CPWA}: \mathcal{X} \rightarrow \mathbb{R}^m$ to denote a Continuous and Piece-Wise Affine (CPWA) function of the form:
\begin{equation}
    \label{eq:cpwa}
    \Psi_\text{CPWA}(x) = K_i x + b_i\quad \text{if}\ x \in \mathcal{R}_i,\ i =1, \ldots, L,
\end{equation}
where the collection of polytopic subsets $\{\mathcal{R}_1, \ldots, \mathcal{R}_L\}$ is a partition of the set $\mathcal{X}$ such that $\bigcup_{i = 1}^L \mathcal{R}_i = \mathcal{X}$ and $\mathrm{Int}(\mathcal{R}_i) \cap \mathrm{Int}(\mathcal{R}_j) = \emptyset$ if $i \neq j$. We call each polytopic subset $\mathcal{R}_i \subset \mathcal{X}$ a linear region, and denote by $\mathbb{L}_{\Psi_\text{CPWA}}$ the set of linear regions associated to $\Psi_\text{CPWA}$, i.e.:
\begin{equation}
    \label{eq:cpwa_lin_reg}
    \mathbb{L}_{\Psi_\text{CPWA}} = \{\mathcal{R}_1, \ldots, \mathcal{R}_L\}.
\end{equation}

\subsection{Dynamical Model, NN Controller, and Specification} 
Consider discrete-time nonlinear dynamical systems of the form:
\begin{equation}
    \label{eq:dyn}
    x^{(t+1)} = f(x^{(t)}, u^{(t)}),
\end{equation}
where the state vector $x^{(t)} \in \mathcal{X} \subset \mathbb{R}^n$, the control vector $u^{(t)} \in \mathcal{U} \subset \mathbb{R}^m$, and $t \in \mathbb{N}$. Given a feedback control law $\Psi: \mathcal{X} \rightarrow \mathcal{U}$, we use $\xi_{x_0, \Psi}: \mathbb{N} \rightarrow \mathcal{X}$ to denote the closed-loop trajectory of~\eqref{eq:dyn} that starts from the state $x_0 \in \mathcal{X}$ and evolves under the control law $\Psi$. 

In this paper, our primary focus is on controlling the nonlinear system~\eqref{eq:dyn} with a state-feedback neural network controller $\mathcal{NN}: \mathcal{X} \rightarrow \mathcal{U}$. An $F$-layer Rectified Linear Unit (ReLU) NN is specified by composing $F$ layer functions (or just layers). A layer $l$ with $\mathfrak{i}_l$ inputs and $\mathfrak{o}_l$ outputs is specified by a weight matrix $W^{(l)} \in \mathbb{R}^{\mathfrak{o}_l \times \mathfrak{i}_l}$ and a bias vector $b^{(l)} \in \mathbb{R}^{\mathfrak{o}_l}$ as follows:
\begin{equation}
    L_{\theta^{(l)}}: z \mapsto \max\{ W^{(l)} z + b^{(l)}, 0 \}, \label{eq:layer_fnc}
\end{equation}
where the $\max$ function is taken element-wise, and $\theta^{(l)} \triangleq (W^{(l)}, b^{(l)})$ for brevity. Thus, an $F$-layer ReLU NN is specified by $F$ layer functions $\{L_{\theta^{(l)}} : l = 1, \dots, F\}$ whose input and output dimensions are composable: that is they satisfy $\mathfrak{i}_{l} = \mathfrak{o}_{l-1}, l = 2, \dots, F$. Specifically:
\begin{equation}
	\mathcal{NN}_\theta(x) = (L_{\theta^{(F)}} \circ L_{\theta^{(F-1)}} \circ \dots \circ L_{\theta^{(1)}})(x),
\end{equation}
where we index a ReLU NN function by a list of matrices $\theta \triangleq (\theta^{(1)}, \dots, \theta^{(F)})$. Also, it is common to allow the final layer function to omit the $\max$ function altogether, and we will be explicit about this when it is the case. 


As a typical control task, this paper considers a reach-avoid specification $\phi$, which combines a safety property $\phi_{\text{safety}}$ for avoiding a set of obstacles $\{\mathcal{O}_1, \ldots, \mathcal{O}_o\}$ with $\mathcal{O}_i \subset \mathcal{X}$, and a liveness property $\phi_{\text{liveness}}$ for reaching a goal region $\mathcal{X}_\text{goal} \subset \mathcal{X}$ in a bounded time horizon $T$. We use $\xi_{x_0, \Psi} \models \phi_{\text{safety}}$ and $\xi_{x_0, \Psi} \models \phi_{\text{liveness}}$ to denote a trajectory $\xi_{x_0, \Psi}$ satisfies the safety and liveness specifications, respectively, i.e.:
\begin{align*}
    &\xi_{x_0, \Psi} \models \phi_{\text{safety}} \Longleftrightarrow \forall t \in \mathbb{N}, \forall i \in \{1, \ldots, o\}, \xi_{x_0, \Psi}(t) \not\in O_i,  \\
    &\xi_{x_0, \Psi} \models \phi_{\text{liveness}} \Longleftrightarrow  \exists	t' \in \{1,\ldots T\}, \; \xi_{x_0, \Psi}(t') \in \mathcal{X}_\text{goal}.
\end{align*}
Given a set of initial states $\mathcal{X}_\text{init}$, a control law $\Psi: \mathcal{X} \rightarrow \mathbb{R}^m$ satisfies the specification $\phi$ (denoted by $\Psi, \mathcal{X}_\text{init} \models \phi$) if all trajectories starting from the set $\mathcal{X}_\text{init}$ satisfy the specification, i.e., $\xi_{x, \Psi} \models \phi$, $\forall x \in \mathcal{X}_\text{init}$.

\subsection{Main Problem} 
Given the dynamical model~\eqref{eq:dyn} and a reach-avoid specification $\phi = \phi_{\text{safety}} \land \phi_{\text{liveness}}$, we consider the problem of designing a NN controller with provable guarantees as described in the next problem.
\begin{prob}
    \label{prob:safe}
    Given the nonlinear dynamical system~\eqref{eq:dyn} and a reach-avoid specification $\phi$, compute 
    an assignment of neural network weights $\theta$ and a set of initial states $\mathcal{X} _\text{init} \subseteq \mathcal{X}$ such that $\mathcal{NN}_\theta, \mathcal{X}_\text{init} \models \phi_{\text{safety}} \land \phi_{\text{liveness}}$.
\end{prob}

While it is desirable to find the \emph{largest} possible set of safe initial states $\mathcal{X} _\text{init}$, our algorithm will instead focus on finding an $\epsilon$ sub-optimal $\mathcal{X} _\text{init}$. For space considerations, the quantification of the sub-optimality in the computations of $\mathcal{X} _\text{init}$ is omitted.


\section{Overview of the Proposed Formal Training Framework} 
\label{sec:framework} 
Before describing our approach to solve Problems~\ref{prob:safe}, we start by recalling the connection between ReLU neural networks and Continuous Piece-Wise Affine (CPWA) functions as follows~\citep{pascanu2013number}:
\begin{prop}
    Every $\mathbb{R}^n \rightarrow \mathbb{R}^m$ ReLU NN represents a continuous piece-wise affine function. 
\end{prop}

In this paper, we confine our attention to CPWA controllers (and hence neural network controllers) that are selected from a bounded polytopic set {$\mathcal{P}^{K} \times \mathcal{P}^{b} \subset \mathbb{R}^{m\times n} \times \mathbb{R}^m$}, i.e., we assume that $K_i \in \mathcal{P}^{K}$ and $b_i \in \mathcal{P}^{b}$.

Our solution to Problem~\ref{prob:safe} is to use the mathematical model of the dynamical system~\eqref{eq:dyn} to 
guide the training of neural networks. 
In particular, our approach is split into two components, one to address the safety specifications while the other one addresses the liveness specifications as described in the next two subsections.

\begin{figure*}[!ht]  
    \center
      \subfloat[]{\includegraphics[height=0.2\linewidth]{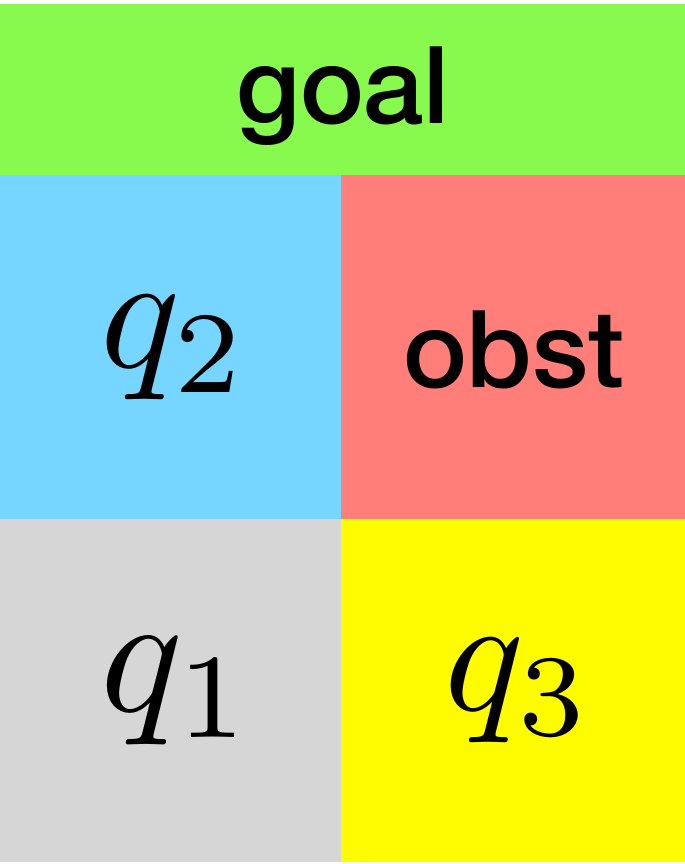}} 
      \hspace{5mm} 
      \subfloat[]{\includegraphics[height=0.2\linewidth]{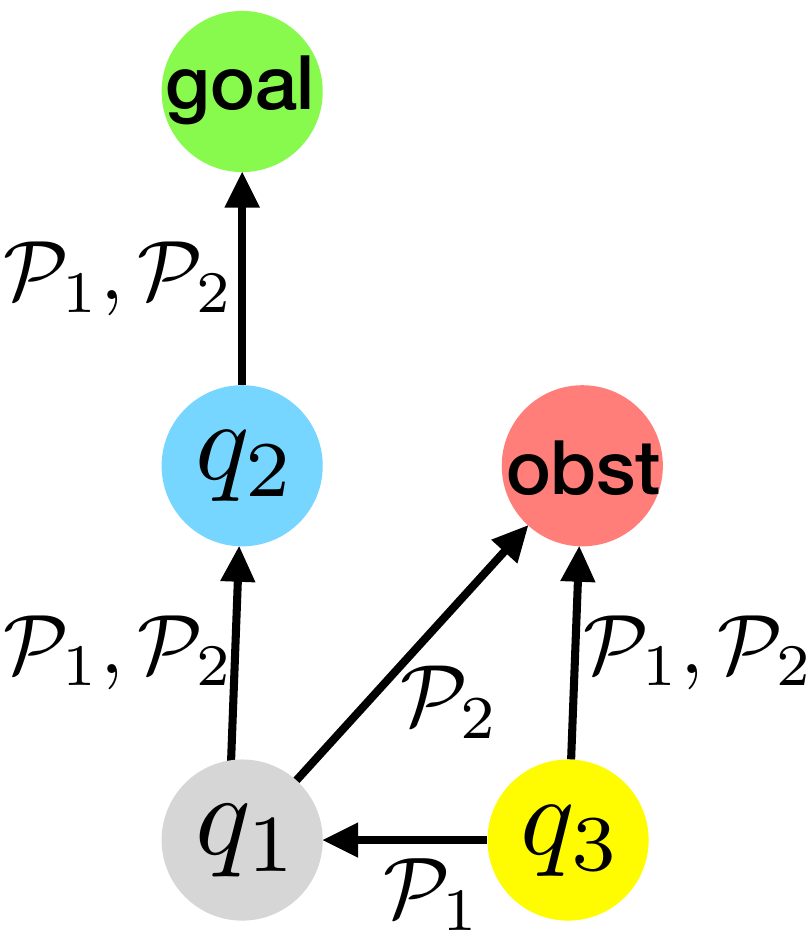}}
      \hspace{5mm}
      \subfloat[]{\includegraphics[height=0.2\linewidth]{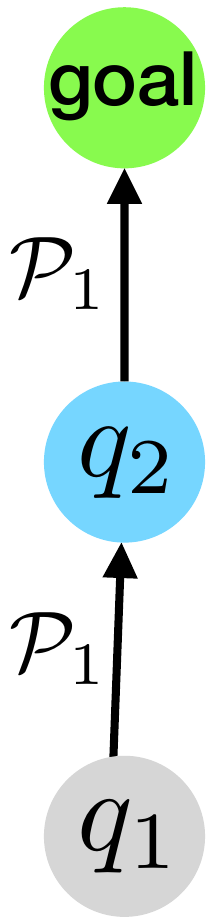}}
      \hspace{5mm}
      \subfloat[]{\includegraphics[height=0.2\linewidth]{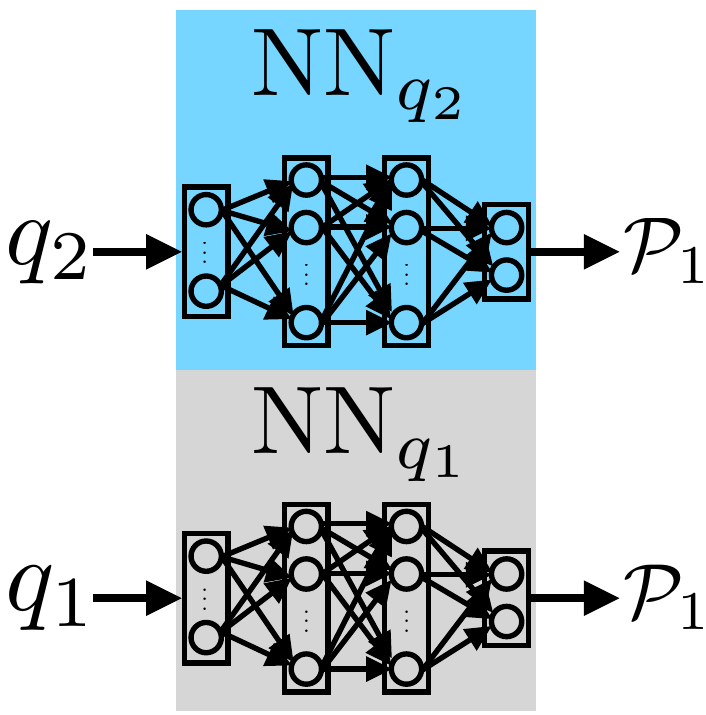}}  
      \hspace{2mm}
      \subfloat[]{\includegraphics[height=0.2\linewidth]{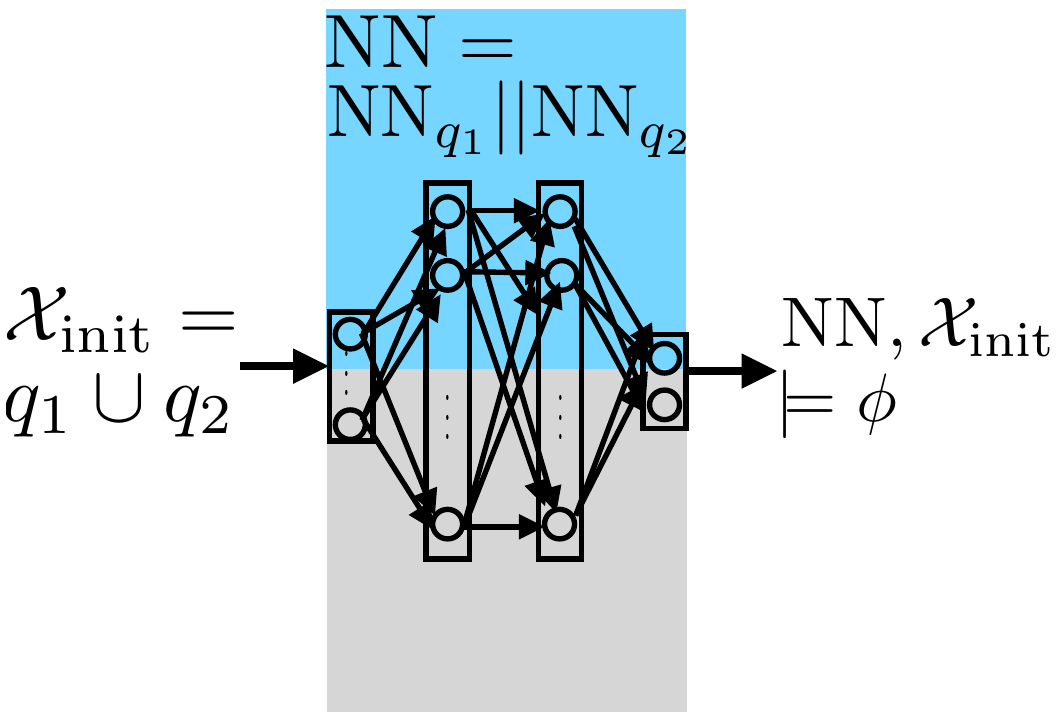}}  
    \caption{(a) An example of a state space is partitioned into abstract states $q_i$, $i=1,2,3$, $\texttt{goal}$ (the goal region), and $\texttt{obst}$ (the obstacle). The controller parameter space is discretized into controller partitions $\mathcal{P}_1$ and $\mathcal{P}_2$. (b) Posterior graph constructed using reachability analysis of the dynamical model.
    (c) Assign one controller partition to each abstract state. Since both the controller partitions $\mathcal{P}_1$ and $\mathcal{P}_2$ are in the label from the state $q_3$ to the obstacle, $q_3$ is unsafe and hence is not considered in (c). (d) Train a local NN for each abstract state and enforce the CPWA functions represented by the NNs are within the assigned controller partitions. (e) Combine local NNs into a single NN controller that is guaranteed to satisfy the given specification $\phi$.} 
    \label{fig:frame} 
  \end{figure*} 

\subsection{Addressing Safety Specification $\phi_{\text{safety}}$} 
Our approach to address the safety specification $\phi_{\text{safety}}$ is as follows:
\begin{itemize}
    \item \textbf{Step 1:} Capture the closed-loop behavior of the system under \emph{all} CPWA controllers using a finite-state abstract model. Such abstract model can be obtained by extending current results on reachability analysis for polytopic systems~\citep{yordanov2012temporal}.
    \item \textbf{Step 2:} Identify a family of CPWA controllers that lead to correct behavior on the abstract model.
    \item \textbf{Step 3:} Enforce the training of the NN to pick from the identified family of the CPWA controllers.
\end{itemize}

Figure~\ref{fig:frame} conceptualizes our framework. To start with, we construct an abstract model by partitioning both the state space $\mathcal{X}$ and the set of all allowed CPWA functions $\mathcal{P}^{K} \times \mathcal{P}^{b}$. In Figure~\ref{fig:frame} (a), the state space $\mathcal{X}$ is partitioned into a set of abstract states $\mathbb{X} = \{q_1, q_2, q_3, \text{obst}, \text{goal}\}$ such that $\mathcal{X} = \bigcup_{q \in \mathbb{X}} q$.
Similarly, the controller space $\mathcal{P}^{K} \times \mathcal{P}^{b}$ is partitioned into a set of controller partitions $\mathbb{P} = \{\mathcal{P}_1, \mathcal{P}_2\}$ such that $\mathcal{P}^{K} \times \mathcal{P}^{b} = \bigcup_{\mathcal{P} \in \mathbb{P}} \mathcal{P}$ (each controller partition $\mathcal{P} \in \mathbb{P}$ represents a subset of CPWA functions). 
The resulting abstract model is a non-deterministic finite transition system with nodes represent abstract states in $\mathbb{X}$ and transitions are labeled by controller partitions in $\mathbb{P}$. Transitions between the abstract states are computed based on the reachable sets of the nonlinear system~\eqref{eq:dyn} from each abstract state $q \in \mathbb{X}$ and under every controller partition $\mathcal{P} \in \mathbb{P}$. 

Based on the abstract model, we compute a function $P_\text{safe}$ that maps each abstract state $q \in \mathbb{X}$ to a subset of the controller partitions
that are considered to be safe at the abstract state $q$, denoted by $P_\text{safe}(q)$. For example, in Figure~\ref{fig:frame} (b), since the transition from $q_1$ labeled by $\mathcal{P}_2$ leads to the obstacle, the controller partition $\mathcal{P}_2$ is unsafe at $q_1$, and hence $P_\text{safe}(q_1) = \{\mathcal{P}_1\}$. Similarly, $P_\text{safe}(q_2) = \{\mathcal{P}_1, \mathcal{P}_2\}$. For the abstract state $q_3$, since both $\mathcal{P}_1$ and $\mathcal{P}_2$ can lead to the obstacle, $P_\text{safe}(q_3)$ is empty, and hence $q_3$ is considered an unsafe abstract state. The set of initial states that can provide safety guarantee is the union of the safe abstract states, i.e., $\mathcal{X}_\text{init} = q_1 \cup q_2$. 

Using the set of safe controllers captured by $P_\text{safe}(q)$, it is direct to show that 
if a neural network gives rise to CPWA functions in $P_\text{safe}(q)$ at every abstract state $q \in \mathbb{X}$, then the NN satisfies the safety specifications. 
Therefore, to ensure the safety of the resulting trained neural network, we propose a NN weight ``projection'' operator to enforce that the trained NN only gives rise to the CPWA functions indicated by $P_\text{safe}(q)$.

\subsection{Addressing Liveness Specification $\phi_{\text{liveness}}$} 
Our approach to addressing the liveness specification $\phi_{\text{liveness}}$ (reaching the goal) can be summarized as follows:
\begin{itemize}
    \item \textbf{Step 1:} Use the abstract model to identify \emph{candidate} controller partitions $\mathcal{P}^\star \in P_\text{safe}(q)$ that can lead to satisfaction of the liveness properties.
    \item \textbf{Step 2:} Train one local neural network $\text{NN}_q$ for each of the abstract states using either imitation learning or reinforcement learning. We take into account the knowledge of $\mathcal{P}^\star$ as constraints during training, and use the NN weight projection operator 
    to ensure that the resulting NN still enjoys the safety specifications.
    \item \textbf{Step 3:} Combine all the local neural networks $\text{NN}_q$ into a single global NN.
\end{itemize}

In the context of the example in Figure~\ref{fig:frame} (c), the controller partition $\mathcal{P}_1$ is assigned to both $q_1$ and $q_2$ as the \emph{candidate} controller partition $\mathcal{P}^\star$ that \emph{may} lead to the satisfaction of the liveness properties.


Next, we train one local neural network $\text{NN}_q$ for each abstract state, or a subset of abstract states assigned with the same controller partition as shown in Figure~\ref{fig:frame} (d). During training of the local neural networks, we project their weights to enforce that the resulting NNs only give rise to the CPWA functions belong to the assigned controller partitions $\mathcal{P}^\star$. Since the controller partition $\mathcal{P}^\star$ is chosen from the subset $P_\text{safe}(q)$ at every $q \in \mathbb{X}$, the resulting NN enjoys the same safety guarantees.



Finally, in Figure~\ref{fig:frame} (e), we combine the local networks trained for each abstract state into a single NN controller, with layers of fixed weights to decide which part of the NN should be activated. 


\subsection{Formal Guarantees}
\label{subsec:guarantee}
We highlight that the proposed framework above \emph{always} guarantees that the resulting NN satisfies the safety specification $\phi_{\text{safety}}$ thanks to the NN weight projection operator. This is reflected in Theorem~\ref{thm:global_safe} in Section~\ref{subsec:combined_nn}.

On the other hand, achieving the liveness specification $\phi_{\text{liveness}}$ depends on the effort of neural network training and the quality of training data, and hence needs an extra step of formally verifying the resulting neural networks and iteratively changing the candidate controller partitions $\mathcal{P}^\star$ if needed. However, we argue that the resulting NN architecture is modular and is composed of a set of local networks $\text{NN}_q$ that are more amenable to verification. The proposed architecture leads to a direct divide-and-conquer approach in which only some of the local networks $\text{NN}_q$ may need to be re-trained whenever the liveness properties are not met.

\section{Provably Safe Training of NN Controllers}
\label{sec:safe_partition}
In this section, we provide details on how to construct the abstract model that captures the behavior of all CPWA controllers along with how to choose the controller partitions that satisfy the safety specifications and train the corresponding NN controllers.

\vspace{-2mm}
\subsection{Step 1: Constructing the Abstract Model} 
\label{subsec:abst_model}
\vspace{-2mm}
In order to capture the behavior of the system~\eqref{eq:dyn} under all possible CPWA controllers $\Psi_\text{CPWA}$ in the form of~\eqref{eq:cpwa}, we construct a finite-state abstract model by discretizing both the state space and the set of all allowed CPWA functions. In particular, we partition the state space $\mathcal{X} \subset \mathbb{R}^n$ into a collection of abstract states, denoted by $\mathbb{X} = \{q_1, \ldots, q_N\}$. The goal region $\mathcal{X}_\text{goal} \subset \mathcal{X}$ is represented by a single abstract state $q_\text{goal} \in \mathbb{X}$, and the set of obstacles $\mathbb{X}_\text{obst} = \bigcup_{i=1}^o \{q_{\text{obst}_i}\}$ represents each obstacle $\mathcal{O}_i \subset \mathcal{X}$ by an abstract state $q_{\text{obst}_i} \in \mathbb{X}$, $i = 1, \ldots, o$. Other abstract states $q_i \in \mathbb{X} \setminus (\mathbb{X}_\text{obst} \cup \{q_\text{goal}\})$ represent infinity-norm balls in the state space $\mathcal{X} \subset \mathbb{R}^n$, and the partitioning of the state space satisfies  $\mathcal{X} = \bigcup_{q \in \mathbb{X}} q$ and $\mathrm{Int}(q_i) \cap \mathrm{Int}(q_j) = \emptyset$ if $i \neq j$. For the interest of this paper, we consider the size of abstract states is pre-specified, by noticing that discretization techniques such as adaptive partitioning (e.g.,~\cite{schmuck18hscc}) can be directly applied in our framework. In experiments, we show results with both uniform and non-uniform partitioning of the state space.

Similarly, we partition the controller space into polytopic subsets. For simplicity of notation, we define the set of parameters $\mathcal{P}^{K \times b} \subset \mathbb{R}^{m \times (n+1)}$ be a polytope that combines $\mathcal{P}^K \in \mathbb{R}^{m \times n}$ and $\mathcal{P}^b \in \mathbb{R}^m$, and with some abuse of notation, we use $K_i(x)$ with a single parameter $K_i \in \mathcal{P}^{K \times b}$ to denote $K'_i x_i + b'_i$ with the pair $(K'_i,b'_i) = K_i$. The controller space  $\mathcal{P}^{K \times b} \subset \mathbb{R}^{m \times (n+1)}$ is discretized into a collection of polytopic subsets in $\mathbb{R}^{m \times (n+1)}$, denoted by $\mathbb{P} = \{\mathcal{P}_1, \ldots, \mathcal{P}_M\}$, such that $\mathcal{P}^{K \times b} = \bigcup_{\mathcal{P} \in \mathbb{P}} \mathcal{P}$ and $\mathrm{Int}(\mathcal{P}_i) \cap \mathrm{Int}(\mathcal{P}_j) = \emptyset$ if $i \neq j$. We call each of the subsets $\mathcal{P}_i \in \mathbb{P}$ a \emph{controller partition}. Each controller partition $\mathcal{P} \in \mathbb{P}$ represents a subset of CPWA functions, by restricting parameters $K_i$ in a CPWA function to take values from $\mathcal{P}$. 


In order to reason about the safety property $\phi_{\text{safety}}$, we introduce a posterior operator based on the dynamical model~\eqref{eq:dyn}. The posterior of an abstract state $q \in \mathbb{X}$ under a controller partition $\mathcal{P} \in \mathbb{P}$ is the set of states that can be reached in one step from states $x \in q$ by using affine state feedback controllers with parameters $K \in \mathcal{P}$, i.e.:
\begin{equation}
    \label{eq:post}
    \mathrm{Post}(q, \mathcal{P}) \triangleq \{f(x, K(x)) \in \mathbb{R}^n \; | \; x \in q,\ K \in \mathcal{P}\}. 
\end{equation}%
Calculating the exact posterior of a nonlinear system is computationally daunting. Therefore, we rely on over-approximations of the posterior set denoted by $\overline{\mathrm{Post}}$.

Our abstract model is defined by using the set of abstract states $\mathbb{X}$, the set of controller partitions $\mathbb{P}$, and the posterior operator $\overline{\mathrm{Post}}$. Intuitively, an abstract state $q \in \mathbb{X}$ has a transition to $q^\prime \in \mathbb{X}$ under a controller partition $\mathcal{P} \in \mathbb{P}$ if the intersection between $q^\prime$ and $\overline{\mathrm{Post}}(q, \mathcal{P})$ is non-empty. 
\begin{defn}
    \label{def:s_post}
    (Posterior Graph)
    A posterior graph is a finite transition system $S_\mathrm{Post} \triangleq (X, X_0, L, \longrightarrow)$, where:
    \begin{itemize}
        \item $X = \mathbb{X}$;
        \item $X_0 = \mathbb{X}$;  
        \item $L = 2^\mathbb{P}$;
        \item $q \overset{l}{\longrightarrow} q^\prime$, if $q \not \in \{q_\text{goal}\} \cup \mathbb{X}_\text{obst}$ and $l = \{\mathcal{P} \in \mathbb{P} \; | \; q^\prime \cap \overline{\mathrm{Post}}(q, \mathcal{P}) \neq \emptyset\} \neq \emptyset$.
    \end{itemize}
\end{defn}%
The computation of the posterior operator can be done by borrowing existing techniques in reachability analysis of polytopic systems~\citep{yordanov2012temporal}, with the subtle difference due to the need to consider polytopic partitions of the parameter space $\mathcal{P}^{K \times b} \subset \mathbb{R}^{m \times (n+1)}$ instead of the well-studied problem of considering polytopic partitions of the input space $\mathcal{U} \subset \mathbb{R}^m$. We refer to the Posterior Graph $S_\mathrm{Post}$ as the finite-state abstract model of~\eqref{eq:dyn}.

\subsection{Step 2: Computing the Function $P_\text{safe}$} 
\label{subsec:compute_p_safe}
Once the abstract model $S_\mathrm{Post}$ is computed, our framework identifies a set of safe controller partitions $P_\text{safe}(q) \subseteq \mathbb{P}$ at each abstract state $q \in \mathbb{X}$. It is possible that the set $P_\text{safe}(q)$ is empty at some state $q \in \mathbb{X}$, in which case, the state $q$ is considered to be unsafe. 

We first introduce the $\mathrm{Next}$ operator as follows:
\begin{equation}
    \label{eq:next}
    \mathrm{Next}(q, \mathcal{P}) \triangleq \{q^\prime \in \mathbb{X} \; | \; q^\prime \cap \overline{\mathrm{Post}}(q, \mathcal{P}) \neq \emptyset\}.
\end{equation}%
We identify the set of unsafe abstract states in a recursive manner by backtracking from the set of obstacles $\mathbb{X}_\text{obst}$ in the posterior graph $S_\mathrm{Post}$:
\vspace{-8mm}
{\small 
\begin{align}
    &\mathbb{X}_\text{unsafe}^0 = \mathbb{X}_\text{obst} \label{eq:x_unsafe^0} \notag \\
    &\mathbb{X}_\text{unsafe}^1 = \{q \in \mathbb{X} \; | \; \forall \mathcal{P} \in \mathbb{P}: \mathrm{Next}(q, \mathcal{P}) \cap \mathbb{X}_\text{unsafe}^0 \neq \emptyset\} \cup \mathbb{X}_\text{unsafe}^0 \notag\\
    &\vdots \notag \\
    &\mathbb{X}_\text{unsafe}^{k} = \{q \in \mathbb{X} \; | \; \forall \mathcal{P} \in \mathbb{P}: \mathrm{Next}(q, \mathcal{P}) \cap \mathbb{X}_\text{unsafe}^{k-1} \neq \emptyset\} \cup \mathbb{X}_\text{unsafe}^{k-1} \notag
\end{align}}%
The backtracking stops at iteration $k^*$ when it cannot find new unsafe states, i.e., $\mathbb{X}_\text{unsafe}^{k^*} = \mathbb{X}_\text{unsafe}^{k^*-1}$. This backtracking process ensures that the set $\mathbb{X}_\text{unsafe}^{k^*}$ includes all the abstract states that inevitably lead to the obstacles $\mathbb{X}_\text{obst}$ (either directly or over multiple transitions).

Then, we define the set of safe abstract states as $\mathbb{X}_\text{safe} \triangleq \mathbb{X} \setminus \mathbb{X}_\text{unsafe}^{k^*}$. The following proposition guarantees that there exists a controller partition assignment for each safe abstract state $q \in \mathbb{X}_\text{safe}$ such that collision-free can be satisfied for all times in the future.
\begin{prop}
    \label{prop:invariant}
    The set of safe abstract states $\mathbb{X}_\text{safe}$ is control positive invariant, i.e.:
    \begin{equation}
        \forall q \in \mathbb{X}_\text{safe},\; \exists \mathcal{P} \in \mathbb{P}\; \text{s.t.}\; \mathrm{Next}(q, \mathcal{P}) \subseteq \mathbb{X}_\text{safe}. \label{eq:invariant}
    \end{equation}
\end{prop}
\vspace{-6mm}
\begin{pf}
    Assume for the sake of contradiction that 
    $\exists q \in \mathbb{X}_\text{safe}$, s.t. $\forall \mathcal{P} \in \mathbb{P}$, $\mathrm{Next}(q, \mathcal{P}) \cap \mathbb{X}_\text{unsafe}^{k^*-1} \neq \emptyset$. Then, by the backtracking process, $q \in \mathbb{X}_\text{unsafe}^{k^*}$ which contradicts the assumption that $q \in \mathbb{X}_\text{safe} = \mathbb{X} \setminus \mathbb{X}_\text{unsafe}^{k^*}$.
\end{pf}
\vspace{-2mm}
Correspondingly, the function $P_\text{safe}$ maps each abstract state $q \in \mathbb{X}_\text{safe}$ to the subset of controller partitions that can be used to force the invariance of the set $\mathbb{X}_\text{safe}$:
\begin{equation}
    \label{eq:compute_p_safe}
    P_{\text{safe}}: q \mapsto \{\mathcal{P} \in \mathbb{P}\ |\ \mathrm{Next}(q, \mathcal{P}) \subseteq \mathbb{X}_\text{safe}\}.
\end{equation}

The following theorem summarizes the safety property.
\begin{thm}
    \label{thm:safe_partition}
    Let $\Psi_{q, \text{CPWA}}: q \rightarrow \mathcal{U}$ be a state feedback CPWA controller defined at an abstract state $q \in \mathbb{X}_\text{safe}$. Assume that the parameters of the affine functions $K_i$ in $\Psi_{q, \text{CPWA}}$ are chosen such that $K_i \in \mathcal{P}$ and $\mathcal{P} \in P_\text{safe}(q)$. Consider the set of initial states $\mathcal{X}_\text{init} = \bigcup_{q \in \mathbb{X}_\text{safe}} q$ and define a controller $\Psi: \mathcal{X}_{\text{safe}} \rightarrow \mathcal{U}$ as:
    \begin{equation}
        \label{eq:Psi}
        \Psi(x) \triangleq
        \begin{cases}
            \Psi_{q_1, \text{CPWA}}(x) \qquad & \forall x \in q_1 \\
            \qquad \vdots \\
            \Psi_{q_{N^\prime}, \text{CPWA}}(x) & \forall x \in q_{N^\prime},
        \end{cases}
    \end{equation}
    where $N^\prime = \vert \mathbb{X}_\text{safe} \vert$ is the number of safe abstract states. 
    Then, it holds that $\Psi, \mathcal{X}_\text{init} \models \phi_\text{safety}.$
\end{thm}   
\vspace{-3mm}
\begin{pf}
    It is sufficient to show that the set $\mathcal{X}_\text{init} = \bigcup_{q \in \mathbb{X}_\text{safe}} q$ is positive invariant under the controller $\Psi$ in the form of~\eqref{eq:Psi}. In other words, for any state $x \in \mathcal{X}_\text{init}$, let $q \in \mathbb{X}_\text{safe}$ be the abstract state contains $x$, i.e., $x \in q$, then by applying any controller $K \in \mathcal{P}$ with any $\mathcal{P} \in P_\text{safe}(q)$, it holds that $f(x, K(x)) \in \mathcal{X}_\text{init}$. 
    
    The definitions of the posterior operator~\eqref{eq:post} and the $\mathrm{Next}$ operator~\eqref{eq:next} directly yield $f(x, K(x)) \in \overline{\mathrm{Post}}(q, \mathcal{P})$ and $\overline{\mathrm{Post}}(q, \mathcal{P}) \subseteq \bigcup_{q^\prime \in \mathrm{Next}(q, \mathcal{P})} q^\prime$, and hence:
    \vspace{-4mm}
    {\small
    \begin{equation}
        \label{eq:pf_subset1}
         f(x, K(x)) \in \bigcup_{q^\prime \in \mathrm{Next}(q, \mathcal{P})} q^\prime,
    \end{equation}}%
    where $x \in q$, $q \in \mathbb{X}$, $K \in \mathcal{P}$, and $\mathcal{P} \in \mathbb{X}$ are arbitrarily chosen. Now consider an arbitrary \emph{safe} abstract state $q \in \mathbb{X}_\text{safe}$, by Proposition~\ref{prop:invariant} and the definition of $P_\text{safe}$ as~\eqref{eq:compute_p_safe}, the set $P_\text{safe}(q) \neq \emptyset$, and for any $\mathcal{P} \in P_\text{safe}(q)$, it holds that:
    \vspace{-4mm}
    {\small
    \begin{equation}
        \label{eq:pf_subset2}
        \bigcup_{q^\prime \in \mathrm{Next}(q, \mathcal{P})} q^\prime \subseteq \mathcal{X}_\text{init},
    \end{equation}}%
    where it uses the definition $\mathcal{X}_\text{init} = \bigcup_{q \in \mathbb{X}_\text{safe}} q$. Therefore, by~\eqref{eq:pf_subset1} and~\eqref{eq:pf_subset2}, $f(x, K(x)) \in \mathcal{X}_\text{init}$ for any $x \in q \in \mathbb{X}_\text{safe}$ and any $K \in \mathcal{P} \in P_\text{safe}(q)$. 
\end{pf}
\vspace{-2mm}
By Theorem~\ref{thm:safe_partition}, the system is guaranteed to satisfy the safety specification $\phi_\text{safety}$ by applying any CPWA controller $\Psi_{q, \text{CPWA}}$ at each abstract sate $q \in \mathbb{X}_\text{safe}$, as long as the controller parameters $K_i$ are chosen from the safe controller partitions $\mathcal{P} \in P_\text{safe}(q)$. This allows us to conclude the safety guarantee provided by a NN controller if the CPWA function represented by the NN is chosen from the safe controller partitions, as we show in detail in the next subsection.

\subsection{Step 3: Safe Training Using NN Weight Projection}
\label{subsec:project_weights}
As described in Section~\ref{sec:framework}, the final NN consists of several local networks $\text{NN}_q$ that will be trained to account for the liveness property. Nevertheless, to ensure that the safety property is met by each of the local networks $\text{NN}_q$, we propose a NN weight projection operator that can be incorporated in the training of these local NNs. Given a controller partition $\mathcal{P}^\star \in P_\text{safe}(q)$, this operator projects the weights of $\text{NN}_q$ to ensure that the network represents a CPWA function belongs to $\mathcal{P}^\star$.

Consider a local network $\text{NN}_q$ of $F$ layers, including $F-1$ hidden layers and an output layer. The NN weight projection operator updates the weight matrix $W^{(F)}$ and the bias vector $b^{(F)}$ of the output layer to $\widehat{W}^{(F)}$ and $\widehat{b}^{(F)}$, respectively, such that the CPWA function represented by the updated local network belongs to the selected controller partition $\mathcal{P}^\star \in P_\text{safe}(q)$ at linear regions intersecting $q$. To be specific, let the CPWA functions represented by the local network $\text{NN}_q$ before and after the weight projection be $\mathcal{NN}_\theta$ and $\mathcal{NN}_{\widehat{\theta}}$, respectively. Then, we formulate the NN weight projection operator as an optimization problem: 
\begin{align}
    &\underset{\widehat{W}^{(F)}, \widehat{b}^{(F)}}{\min}\; \underset{x \in q}{\max}\; |\!|{\mathcal{NN}_{\widehat{\theta}}(x) - \mathcal{NN}_\theta(x)}|\!|_1 \label{eq:proj_prototype_obj} \\
    &\text{s.t.}\; K_i \in \mathcal{P}^\star,\; \forall \mathcal{R}_i \in \{\mathcal{R} \in \mathbb{L}_{\mathcal{NN}_\theta} | \mathcal{R} \cap q \neq \emptyset\} \label{eq:proj_prototype_const}
\end{align}%
where by the notation~\eqref{eq:cpwa_lin_reg}, $\mathbb{L}_{\mathcal{NN}_\theta} = \{\mathcal{R}_1, \ldots, \mathcal{R}_L\}$ is the set of linear regions of the CPWA function $\mathcal{NN}_\theta$. Consider the CPWA function $\mathcal{NN}_{\widehat{\theta}}$ is in the form of~\eqref{eq:cpwa}, where affine functions parametrized by $K_i \in \mathbb{R}^{m \times (n+1)}$ are associated to linear regions $\mathcal{R}_i$, $i = 1, \ldots, L$. Then, the constraints~\eqref{eq:proj_prototype_const} require that $K_i \in \mathcal{P}^\star$ whenever the corresponding linear region $\mathcal{R}_i$ intersects the abstract state $q$. As a common practice, we consider there is no ReLU activation function in the output layer. Then, the projection operator does not change the set of linear regions, i.e., $\mathbb{L}_{\mathcal{NN}_{\widehat{\theta}}} = \mathbb{L}_{\mathcal{NN}_\theta}$, since the linear regions of a neural network only depend on the hidden layers, while the projection operator updates the output layer weights.

The optimization problem~\eqref{eq:proj_prototype_obj}-\eqref{eq:proj_prototype_const} tries to minimize the change of NN outputs due to the projection, which is measured by the largest 1-norm difference between the control signals $\mathcal{NN}_{\widehat{\theta}}$ and $\mathcal{NN}_\theta$ across the abstract state $q$, i.e., $\underset{x \in q}{\max}\; |\!|{\mathcal{NN}_{\widehat{\theta}}(x) - \mathcal{NN}_\theta(x)}|\!|_1$. The relation between this objective function and the change of output layer weights is captured in the following proposition.
\begin{prop} \label{prop:weight_relation}
    Consider an $F$-layer local network $\text{NN}_q$ whose output layer weights change from $W^{(F)}$, $b^{(F)}$ to $\widehat{W}^{(F)}$, $\widehat{b}^{(F)}$. Then, the largest change of the NN's outputs across the abstract state $q$ is bounded as follows:
    \vspace{-5mm}
    {\small
    \begin{align}
        &\underset{x \in q}{\max}\; |\!|{\mathcal{NN}_{\widehat{\theta}}(x) - \mathcal{NN}_\theta(x)}|\!|_1 \label{eq:ub_output_diff} \\
        &\leq \underset{x \in \mathrm{Vert}(\mathbb{L}_{\mathcal{NN}_\theta \cap q})}{\max}\; \sum_{i=1}^{m} \sum_{j=1}^{\mathfrak{o}_{F-1}} |\Delta W_{ij}^{(F)}| h_j(x) + \sum_{i=1}^{m} |\Delta b_i^{(F)}| \notag
    \end{align}}%
    where $\Delta W_{ij}^{(F)}$ and $\Delta b_i^{(F)}$ are the $(i,j)$-th and the $i$-th entry of $\Delta W^{(F)} = \widehat{W}^{(F)} - W^{(F)}$ and $\Delta b^{(F)} = \widehat{b}^{(F)} - b^{(F)}$, respectively. 
\end{prop}%
In the above proposition, by following the notation of layer functions~\eqref{eq:layer_fnc}, we use a single function $h: \mathbb{R}^n \rightarrow \mathbb{R}^{\mathfrak{o}_{F-1}}$ to represent all the hidden layers:
\begin{equation}
    h(x) = (L_{\theta^{(F-1)}} \circ L_{\theta^{(F-2)}} \circ \dots \circ L_{\theta^{(1)}})(x), \label{eq:h_hidden}
\end{equation}%
where $\mathfrak{o}_{F-1}$ is the number of neurons in the $(F-1)$-layer (the last hidden layer). We denote by $\mathbb{L}_{\mathcal{NN}_\theta \cap q}$ the intersected regions between the linear regions in $\mathbb{L}_{\mathcal{NN}_\theta}$ and the abstract state $q$, i.e.:
\begin{equation}
    \mathbb{L}_{\mathcal{NN}_\theta \cap q} \triangleq \{\mathcal{R} \cap q\; |\; \mathcal{R} \in \mathbb{L}_{\mathcal{NN}_\theta}, \mathcal{R} \cap q \neq \emptyset\},
\end{equation}%
and denote by $\mathrm{Vert}(\mathbb{L}_{\mathcal{NN}_\theta \cap q})$ the set of all vertices of regions in $\mathbb{L}_{\mathcal{NN}_\theta \cap q}$, i.e., $\mathrm{Vert}(\mathbb{L}_{\mathcal{NN}_\theta \cap q}) \triangleq \bigcup_{\mathcal{R} \in \mathbb{L}_{\mathcal{NN}_\theta \cap q}}\ \mathrm{Vert}(\mathcal{R})$. 
\begin{pf}
    The function $\mathcal{NN}_\theta$ can be written as $\mathcal{NN}_\theta(x) = W^{(F)} h(x) + b^{(F)}$, and after the change of output layer weights, $\mathcal{NN}_{\widehat{\theta}}(x) = \widehat{W}^{(F)} h(x) + \widehat{b}^{(F)}$.~Then, 
    \vspace{-6mm}
    {\small
    \begin{align}
        &\hspace{-3.9pt}\underset{x \in q}{\max}\; |\!|{\mathcal{NN}_{\widehat{\theta}}(x) - \mathcal{NN}_\theta(x)}|\!|_1  \\
        &\hspace{-3.9pt}= \underset{x \in q}{\max}\; \sum_{i=1}^{m} |\sum_{j=1}^{\mathfrak{o}_{F-1}} \Delta W_{ij}^{(F)} h_j(x) + \Delta b_i^{(F)}| \label{eq:ineq1}  \\
        &\hspace{-3.9pt}\leq \underset{x \in q}{\max}\; \sum_{i=1}^{m} \sum_{j=1}^{\mathfrak{o}_{F-1}} |\Delta W_{ij}^{(F)}| h_j(x) + \sum_{i=1}^{m} |\Delta b_i^{(F)}| \label{eq:ineq2} \\
        &\hspace{-3.9pt}=\underset{x \in \mathrm{Vert}(\mathbb{L}_{\mathcal{NN}_\theta \cap q})}{\max} \sum_{i=1}^{m} \sum_{j=1}^{\mathfrak{o}_{F-1}} |\Delta W_{ij}^{(F)}| h_j(x) + \sum_{i=1}^{m} |\Delta b_i^{(F)}| \label{eq:ineq3}
    \end{align}}%
    where~\eqref{eq:ineq1} directly follows the form of $\mathcal{NN}_\theta$ and $\mathcal{NN}_{\widehat{\theta}}$,~\eqref{eq:ineq2} takes the absolute value of each term before summation and uses the fact that the hidden layer outputs $h(x) \geq 0$. We notice that when $h$ is restricted to each linear region of $\mathcal{NN}_\theta$,~\eqref{eq:ineq2} is a linear program whose optimal solution is attained at extreme points. Therefore, in~\eqref{eq:ineq3}, the maximum can be taken over a \emph{finite} set of all the vertices of linear regions in $\mathbb{L}_{\mathcal{NN}_\theta \cap q}$.
\end{pf}




Proposition~\ref{prop:weight_relation} proposes a direct way to design the intended projection operator. It implies that minimizing the right hand side of the inequality~\eqref{eq:ub_output_diff} will minimize the upper bound on the change in the control signal due to the projection operator. To that end, given a controller partition $\mathcal{P}^\star \in P_\text{safe}(q)$, let $\Pi_{\mathcal{P}^\star}$ be the NN weight projection operator that updates the output layer weights of $\text{NN}_q$ as $\widehat{W}^{(F)}, \widehat{b}^{(F)} = \Pi_{\mathcal{P}^\star}(\mathcal{NN}_\theta)$. In order to minimize the change of NN outputs due to the projection, we minimize its upper bound in~\eqref{eq:ub_output_diff}. Accordingly, we write the NN weight projection operator $\Pi_{\mathcal{P}^\star}$ as the following optimization problem:
\vspace{-3mm}
{\small
\begin{align}
    &\underset{\widehat{W}^{(F)}, \widehat{b}^{(F)}}{\text{argmin}}\; \underset{x \in \mathrm{Vert}(\mathbb{L}_{\mathcal{NN}_\theta \cap q})}{\max}\ \sum_{i=1}^{m} \sum_{j=1}^{\mathfrak{o}_{F-1}} |\Delta W_{ij}^{(F)}| h_j(x) + \sum_{i=1}^{m} |\Delta b_i^{(F)}| \label{eq:proj_operator_obj} \\
    &\text{s.t.}\; K_i \in \mathcal{P}^\star,\; \forall \mathcal{R}_i \in \{\mathcal{R} \in \mathbb{L}_{\mathcal{NN}_\theta} | \mathcal{R} \cap q \neq \emptyset\}.  \label{eq:proj_operator_const}
\end{align}}%

A direct question is related to the computational complexity of the proposed projection operator $\Pi_{\mathcal{P}^\star}$. This is addressed in the following proposition.
\begin{prop}
    The NN weight projection operator $\Pi_{\mathcal{P}^\star}$ defined as the optimization problem~\eqref{eq:proj_operator_obj}-\eqref{eq:proj_operator_const} is a linear program.
\end{prop}
\begin{pf}
    Using the epigraph form of the problem~\eqref{eq:proj_operator_obj}-\eqref{eq:proj_operator_const}, we can write it in the equivalent form:
    \vspace{-6mm}
    {\small
    \begin{align}
        &\underset{\widehat{W}^{(F)}, \widehat{b}^{(F)}, t, s_{ij}, v_i}{\min}\; t \qquad \text{such that} \notag\\
        &\sum_{i=1}^{m} \sum_{j=1}^{\mathfrak{o}_{F-1}} s_{ij} h_j(x) + \sum_{i=1}^{m} v_i \leq t,\; \forall x \in \mathrm{Vert}(\mathbb{L}_{\mathcal{NN}_\theta \cap q}) \label{eq:const_t} \\
        &|\widehat{W}_{ij}^{(F)} - W_{ij}^{(F)}| \leq s_{ij},\; i =1, \ldots, m,\; j =1, \ldots, \mathfrak{o}_{F-1}\\
        &|\widehat{b}_i^{(F)} - b_i^{(F)}| \leq v_i,\; i =1, \ldots, m\\
        &K_i \in \mathcal{P}^\star,\; \forall \mathcal{R}_i \in \{\mathcal{R} \in \mathbb{L}_{\mathcal{NN}_\theta}\; |\; \mathcal{R} \cap q \neq \emptyset\}. \label{eq:const_p}
    \end{align}}%
    The inequality~\eqref{eq:const_t} is affine since the hidden layer function $h$ is known and does not depend on the optimization variables. The number of inequalities in~\eqref{eq:const_t} is finite since the set of vertices $\mathrm{Vert}(\mathbb{L}_{\mathcal{NN}_\theta \cap q})$ is finite. To see the constraint~\eqref{eq:const_p} is affine, consider the CPWA function $\mathcal{NN}_{\widehat{\theta}}: x \mapsto \widehat{W}^{(F)} h(x) + \widehat{b}^{(F)}$, which is represented by the NN with the output layer weights $\widehat{W}^{(F)}$, $\widehat{b}^{(F)}$, and the hidden layer function $h$. Since $h$ restricted to each linear region $\mathcal{R}_i \in \mathbb{L}_{\mathcal{NN}_\theta}$ is a known affine function, the parameter $K_i$ (coefficients of $\mathcal{NN}_{\widehat{\theta}}$ restricted to $\mathcal{R}_i$) affinely depends on $\widehat{W}^{(F)}$ and $\widehat{b}^{(F)}$. 
\end{pf}

Finally, the following result shows the safety guarantee provided by applying the projection operator $\Pi_{\mathcal{P}^\star}$ during training of the neural network controller.
\begin{thm}
    \label{thm:project_weights}
    Given an abstract state $q \in \mathbb{X}_\text{safe}$ and an arbitrary controller partition $\mathcal{P}^\star \in P_\text{safe}(q)$. Let the weights of $\text{NN}_q$ be projected by the operator $\Pi_{\mathcal{P}^\star}$ (i.e., the optimization problem~\eqref{eq:proj_operator_obj}-\eqref{eq:proj_operator_const}), then $\text{NN}_q$ is guaranteed to be safe at $q$, i.e., $\text{NN}_q, q \models \phi_{\text{safety}}$.
\end{thm}
\begin{pf}
    The proof directly follows Theorem~\ref{thm:safe_partition} and the constraints~\eqref{eq:proj_operator_const} in the NN weight projection operator. 
\end{pf}

\subsection{Safety-biased Training to Reduce the Effect of Projection}
As a practical issue, it is often desirable to have the trained NNs be safe or close to safe even before the weight projection, and thus the projection operator $\Pi_{\mathcal{P}^\star}$ does not or only slightly modifies the NN weights without compromising performance. To reduce the effect of projection, our algorithm alternates the training and projection, and uses the safe controller partitions $\mathcal{P}^\star \in P_\text{safe}(q)$ as constraints to bias the training. 

We show the provably safe training of each local network $\text{NN}_q$ in Algorithm~\ref{alg:safe_train}. Similar to a projected gradient algorithm, Algorithm~\ref{alg:safe_train} alternates the training (Line~\ref{line:contrained_train} in Algorithm~\ref{alg:safe_train}) and the NN weight projection (Line~\ref{line:project} in Algorithm~\ref{alg:safe_train}) up to a pre-specified maximum iteration $\text{max\_iter}$. The training approach $\texttt{Constrained-Train}$ (Line~\ref{line:contrained_train} in Algorithm~\ref{alg:safe_train}) can be any imitation learning or reinforcement learning algorithm, with some extra constraints required by the safe controller partition $\mathcal{P}^\star \in P_\text{safe}(q)$. In the imitation learning setting, we use $\{(x, u)\}$ to denote the collection of training data generated by an expert, where each state $x \in \mathcal{X}$ is associated with an input label $u \in \mathbb{R}^m$. The approach $\texttt{Constrained-Train}$ requires that all the training data satisfy $u = K(x)$ and $K \in \mathcal{P}^\star$. Similarly, in the reinforcement learning setting, we assign high reward to the explored state-action pairs $(x, u)$ that satisfy $u = K(x)$ and $K \in \mathcal{P}^\star$. 

To guarantee that the trained neural networks are safe, Algorithm~\ref{alg:safe_train} identifies the set of linear regions $\mathbb{L}_{\mathcal{NN}_\theta}$ (Line~\ref{line:identify_LR} in Algorithm~\ref{alg:safe_train}) and projects the neural network weights by solving the linear program~\eqref{eq:proj_operator_obj}-\eqref{eq:proj_operator_const} (Line~\ref{line:project} in Algorithm~\ref{alg:safe_train}). The solved weights $\widehat{W}^{(F)}$, $\widehat{b}^{(F)}$ are then used to replace the currently trained output layer weights $W^{(F)}$, $b^{(F)}$ (Line~\ref{line:update} in Algorithm~\ref{alg:safe_train}), and the largest possible change of the NN's outputs is bounded by~\eqref{eq:ub_output_diff}. Thanks to the training approach $\texttt{Constrained-Train}$ takes the constraints required by $\mathcal{P}^\star$ into consideration, and the problem~\eqref{eq:proj_operator_obj}-\eqref{eq:proj_operator_const} tries to minimize the change of NN outputs due to the projection, we observe in our experiments that the projection operator usually does not change the trained NN weights much. Indeed, in the result section, we show that the trained NNs perform well even with just a single projection at the end (i.e., $\text{max\_iter} = 1$). 

\begin{algorithm}[!t]
    \caption{\textsc{SAFE-TRAIN} ($q, \mathcal{P}^\star$)}
    \label{alg:safe_train}
    {\small
    \begin{algorithmic}[1]
        \State Initialize local network $\text{NN}_q$, $i=1$ 
        \For{$i \leq \text{max\_iter}$} \label{line:max_iter}
            \State $\text{NN}_q = \texttt{Constrained-Train}(\text{NN}_q, \mathcal{P}^\star)$ \label{line:contrained_train}
            \State $\mathbb{L}_{\mathcal{NN}_\theta} = \texttt{Identify-LR}(\text{NN}_q)$ \label{line:identify_LR}
            \State $\widehat{W}^{(F)}, \widehat{b}^{(F)} = \Pi_{\mathcal{P}^\star} (\mathcal{NN}_\theta)$ \label{line:project}
            \State Set the output layer weights of $\text{NN}_q$ be $\widehat{W}^{(F)}, \widehat{b}^{(F)}$ \label{line:update}
            \EndFor
        \State \textbf{Return} $\text{NN}_q$ 
    \end{algorithmic}  
    }
\end{algorithm}

\section{Extension to Liveness Property} \label{sec:liveness}  
\subsection{Controller Partition Assignment} 
\label{subsec:assign_partition}
Among all the safe controller partitions in $P_\text{safe}(q)$, we assign one of them $\mathcal{P}^\star \in P_\text{safe}(q)$ to each abstract state $q \in \mathbb{X}_\text{safe}$, by taking into account the liveness specification $\phi_{\text{liveness}}$. The liveness property requires that the nonlinear system~\eqref{eq:dyn} can reach the goal $\mathcal{X}_\text{goal} \subset \mathcal{X}$ by using the trained NN controller. Unlike the safety property which can be enforced using the projection operator $\Pi_{\mathcal{P}^\star}$, achieving the liveness property depends on practical issues, such as the amount of training data and the effort spent on training the NNs. To that end, we show that the safety guarantee provided by our algorithm does not impede training NNs to satisfy the liveness property by using standard learning techniques. 

Since the posterior graph $S_\mathrm{Post}$ over-approximates the behavior of the system, a transition from the abstract state $q$ to $q^\prime$ in $S_\mathrm{Post}$ does not guarantee that every state $x \in q$ can reach $q^\prime$ in one step. Therefore, we introduce a predecessor operator to capture the liveness property. The predecessor of an abstract state $q^\prime \in \mathbb{X}$ under a controller partition $\mathcal{P} \in \mathbb{P}$ is defined as the set of states that can reach $q^\prime$ in one step by using an affine state feedback controller with some parameter $K \in \mathcal{P}$:
\begin{equation*}
    \mathrm{Pre}(q^\prime, \mathcal{P}) \triangleq \{x \in \mathbb{R}^n \; | \; \exists K \in \mathcal{P}: \; f(x, K(x)) \in q^\prime\}.
\end{equation*}

The predecessor operator can be computed using backward reachability analysis of polytopic systems~\citep{yordanov2012temporal}. We then introduce the predecessor graph:
\begin{defn}
    (Predecessor Graph)
    A predecessor graph is a finite transition system $S_\mathrm{Pre} \triangleq (X, X_0, L, \longrightarrow)$, where:
    \begin{itemize}
        \item $X = \mathbb{X}_\text{safe} \cup \{q_\text{goal}\}$;
        \item $X_0 = \mathbb{X}_\text{safe}$;  
        \item $L = 2^\mathbb{P}$;
        \item $q \overset{l}{\longrightarrow} q^\prime$, if $q \neq q_\text{goal}$ and $l = \{\mathcal{P} \in P_\text{safe}(q)\ |\ q \cap \mathrm{Pre}(q^\prime, \mathcal{P}) \neq \emptyset\} \neq \emptyset$.
    \end{itemize}
\end{defn}
Notice that in the construction of $S_\mathrm{Pre}$, we restrict transition labels to the safe controller partitions $\mathcal{P} \in P_\text{safe}(q)$ at each state $q \in \mathbb{X}_\text{safe}$. Let $\mathcal{T}_\mathrm{Pre}$ be the set of all trajectories over the predecessor graph $S_\mathrm{Pre}$, then we use $\pi_X^{(t)}: \omega \mapsto q$ to denote the map from a trajectory $\omega \in \mathcal{T}_\mathrm{Pre}$ to the abstract state $q$ at time step $t$ in $\omega$, and use $\pi_L^{(t)}: \omega \mapsto l$ to denote the map from $\omega \in \mathcal{T}_\mathrm{Pre}$ to the label $l \in 2^\mathbb{P}$ associated to the transition from $\pi_X^{(t)}(\omega)$ to $\pi_X^{(t+1)}(\omega)$. By extending the formulation of specification to the abstract state space, a trajectory $\omega$ in $S_\mathrm{Pre}$ satisfies the reach-avoid specification $\phi$, denoted by $\omega \models \phi$, if $\omega$ reaches the goal state $q_\text{goal}$ in $T$ steps.
Similar to the notation introduced in the posterior graph, we define a $\mathrm{Next}$ operator in the predecessor graph:
\begin{equation*}
    \mathrm{Next}(q, \mathcal{P}) \triangleq \{q^\prime \in \mathbb{X}_\text{safe} \cup \{q_\text{goal}\}\; |\; q \cap \mathrm{Pre}(q^\prime, \mathcal{P}) \neq \emptyset\}.
\end{equation*}
 
At each abstract state $q \in \mathbb{X}_\text{safe}$, our objective is to choose the candidate controller partition $\mathcal{P}^\star \in P_\text{safe}(q)$ that can lead most of the states $x \in q$ to the goal. To that end, we restrict our attention to the trajectories in $S_\mathrm{Pre}$ that progress towards the goal. That is, let $|\omega|$ be the length of a trajectory $\omega \in \mathcal{T}_\mathrm{Pre}$, and $\mathrm{Dist}: \mathbb{X}_\text{safe} \rightarrow \mathbb{N}$ map a state $q \in \mathbb{X}_\text{safe}$ to the length of the shortest trajectory from the state $q$ to the goal in the predecessor graph $S_\mathrm{Pre}$. Then, we use $\mathcal{T}^\prime_\mathrm{Pre} \subseteq \mathcal{T}_\mathrm{Pre}$ to denote the subset of trajectories that lead to the goal:
\begin{align}
    \mathcal{T}^\prime_\mathrm{Pre} \triangleq &\{\omega \in \mathcal{T}_\mathrm{Pre}\; |\; \mathrm{Dist}(\pi_X^{(t)}(\omega)) < \mathrm{Dist}(\pi_X^{(t-1)}(\omega)), \notag \\ 
    &t = 1, \ldots, |\omega|-1\}. 
\end{align}%
Now we can define the subset of abstract states that progress towards the goal from abstract state $q$ under controller partition $\mathcal{P}$, denoted by $\mathcal{Q}_{q, \mathcal{P}} \subseteq \mathrm{Next}(q, \mathcal{P})$, as the set of abstract states along trajectories $\omega \in \mathcal{T}^\prime_\mathrm{Pre}$ that satisfy the given specification:
\begin{align} \label{eq:Qp}
    \mathcal{Q}_{q, \mathcal{P}} \triangleq &\{q^\prime \in \mathrm{Next}(q, \mathcal{P})\; |\; \exists \omega \in \mathcal{T}^\prime_\mathrm{Pre}: \omega \models \phi, \notag \\
    &\pi_X^{(0)}(\omega)=q, \pi_X^{(1)}(\omega)=q^\prime, \mathcal{P} \in \pi_L^{(0)}(\omega)\}.
\end{align}
Then, the intersection between the abstract state $q$ and the predecessors of abstract states $q^\prime \in \mathcal{Q}_{q, \mathcal{P}}$ under the controller partition $\mathcal{P}$, i.e.:
\begin{equation}
    \mathcal{I}_{q, \mathcal{P}} \triangleq \bigcup_{q^\prime \in \mathcal{Q}_{q, \mathcal{P}}} \left(q \cap \mathrm{Pre}(q^\prime, \mathcal{P})\right),
\end{equation}
measures the portion of states $x \in q$ that progress towards the goal by applying input $u = K(x)$ with $K \in \mathcal{P}$ at the current step. Then, our algorithm assigns each abstract state $q \in \mathbb{X}_\text{safe}$ with the controller partition $\mathcal{P}^\star \in P_\text{safe}(q)$ corresponding to the maximum volume of $\mathcal{I}_{q, \mathcal{P}}$, i.e. $\mathcal{P}^\star = \underset{\mathcal{P} \in P_\text{safe}(q)}{\text{argmax}} \mu(\mathcal{I}_{q, \mathcal{P}})$, where $\mu$ is the Lebesgue measure of $\mathbb{R}^n$.
Indeed, and as mentioned in Section~\ref{subsec:guarantee}, such procedure only ranks the available choices of controller partitions and one may need to iterate over the remaining choices using the same heuristic above.

\subsection{Combined NN Controller}
\label{subsec:combined_nn}
\begin{figure}
    \center
    \subfloat[]{\includegraphics[width=0.23\textwidth]{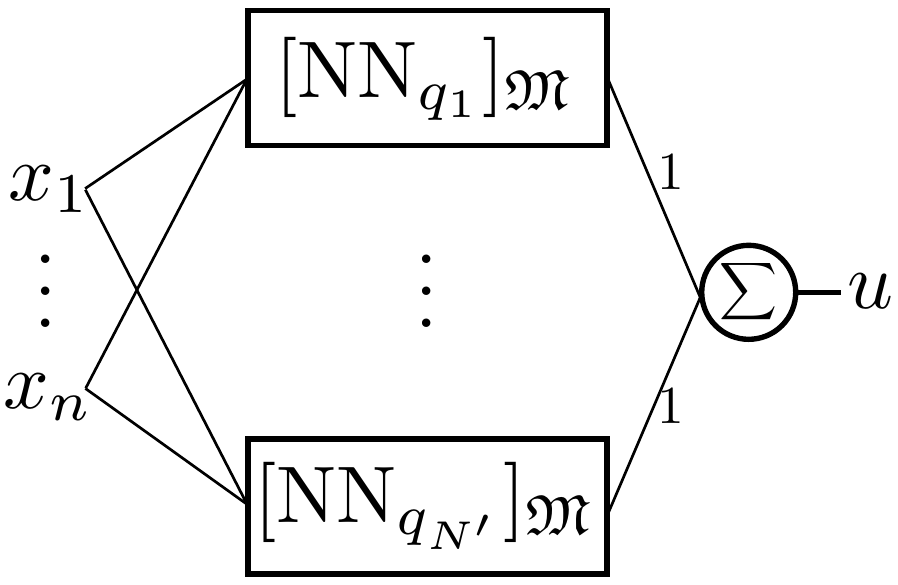}} 
    \subfloat[]{\includegraphics[width=0.25\textwidth]{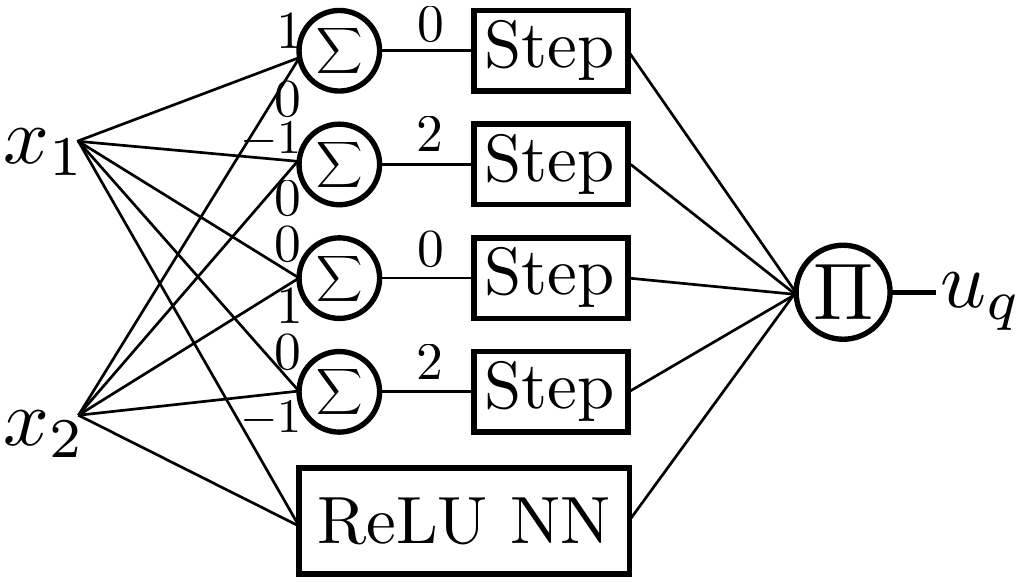}}
    \caption{(a) The global NN controller consists of one module $[\text{NN}_{q_i}]_\mathfrak{M}$ for each abstract state $q_i \in \mathbb{X}_\text{safe}$. (b) An example of the module $[\text{NN}_q]_\mathfrak{M}$, where the state $q \subset \mathbb{R}^2$ is given by $q = [0,2] \times [0,2]$.}
    \label{fig:arch} 
\end{figure}

The final step of our framework is to combine all the local networks $\text{NN}_q$ into one global NN controller.
Figure~\ref{fig:arch} (a) shows the overall structure of the global NN controller obtained by combining modules $[\text{NN}_q]_\mathfrak{M}$ corresponding to the local networks $\text{NN}_q$. As input to the NN controller, the state $x \in \mathcal{X}$ is fed into all the local networks, and the output of the NN controller is the summation of all the local network outputs. In the figures, we show a single output for simplicity, but it can be easily extended to multiple outputs (indeed, even in the figures, $u$ and $u_q$ can be thought as vectors in $\mathbb{R}^m$, and the summation and product operations correspond to vector addition and scalar multiplication, respectively).  

Each module $[\text{NN}_q]_\mathfrak{M}$ consists of two parts: logic and ReLU NN. The logic component decides whether the current state $x$ is in the abstract state $q$ associated with $[\text{NN}_q]_\mathfrak{M}$, and outputs $1$ if the answer is affirmative, $0$ otherwise. The ReLU network $\text{NN}_q$ is trained for this abstract state $q$ using Algorithm~\ref{alg:safe_train}. By multiplying the outputs of the logic and the ReLU NN, output of the module $[\text{NN}_q]_\mathfrak{M}$ is identical to the output of the ReLU NN if $x \in q$, and zero otherwise. Figure~\ref{fig:arch} (b) is an example of the module $[\text{NN}_q]_\mathfrak{M}$ for an arbitrary abstract state $q \subset \mathbb{R}^2$ given by $0 \leq x_1 \leq 2$ and $0 \leq x_2 \leq 2$. 

The logic component in each module $[\text{NN}_q]_\mathfrak{M}$ can be implemented as a single layer NN with fixed weights. Given the $H$-representation $Ax \leq c$ of the abstract state $q$, the weight matrix and the bias vector of the single layer NN are $W^{(1)} = -A$ and $b^{(1)} = c$, respectively. Essentially, this choice of weights encodes each hyperplane inequality in the $H$-representation to each neuron in the single layer. To represent whether an inequality holds, we use a step function as the nonlinear activation function for the single layer: $\mathrm{Step}(x) = 1 \text{ if } x \geq 0$, $0$ otherwise. 
The product of the outputs of all the neurons in the single layer is computed at the end (by the product operator $\Pi$), and hence the logic component returns $1$ if and only if all the hyperplane inequalities are satisfied. 

We refer to the single layer NN for the logic component as $\text{NN}_{q, \Pi}$, along with the ReLU network $\text{NN}_q$, each module can be written as $[\text{NN}_q]_\mathfrak{M} = \text{NN}_q || \text{NN}_{q, \Pi}$, where $||$ denotes the parallel composition. Using the same notation, we can write the global NN controller as:
\begin{equation}
    \text{NN} = [\text{NN}_{q_1}]_\mathfrak{M}\; ||\; \ldots\; ||\; [\text{NN}_{q_{N^\prime}}]_\mathfrak{M}, \label{eq:global_nn}
\end{equation}
where $N^\prime = |\mathbb{X}_\text{safe}|$ is the number of safe abstract states. We summarize the guarantees on the global NN controller in Theorem~\ref{thm:global_safe}, whose proof follows directly from the discussion above along with Theorem~\ref{thm:safe_partition} and Theorem~\ref{thm:project_weights}. Let $\mathcal{Q}_{q_i, \mathcal{P}^\star_i}$ be defined as~\eqref{eq:Qp} and $\mathrm{Reach}(q, \text{NN}_{q})$ be the set of states that can be reached in one step from abstract state $q$ using local network $\text{NN}_q$:
\begin{equation}
\mathrm{Reach}(q, \text{NN}_q) \triangleq \{f(x, \text{NN}_q(x))\ |\ x \in q\}.
\end{equation} 
\vspace{-6mm}
\begin{thm}
    \label{thm:global_safe}
    Consider the nonlinear system~\eqref{eq:dyn} and the reach-avoid specification $\phi = \phi_{\text{safety}} \land \phi_{\text{liveness}}$. Let the controller partition assignment $\mathcal{P}^{\star}_1, \ldots, \mathcal{P}^{\star}_{N^\prime}$ and the local neural networks $\text{NN}_{q_1}, \ldots, \text{NN}_{q_{N^\prime}}$ satisfy (i) $\mathcal{P}^{\star}_i \in P_{\text{safe}}(q_i)$ and (ii) the neural network weights $\theta_i$ of $\text{NN}_{q_i}$ is projected on $\mathcal{P}^{\star}_i$ using the projection operator~\eqref{eq:proj_operator_obj}-\eqref{eq:proj_operator_const}. Then, the global neural network controller $\text{NN}$ given by~\eqref{eq:global_nn} is guaranteed to be safe, i.e., $\text{NN}, \mathcal{X}_\text{init} \models \phi_{\text{safety}}$ with $\mathcal{X}_\text{init} = \bigcup_{q \in \mathbb{X}_\text{safe}} q$. Moreover, if all the local networks $\text{NN}_{q_i}$ satisfy $\mathrm{Reach}(q_i, \text{NN}_{q_i}) \subseteq \bigcup_{q^\prime \in \mathcal{Q}_{q_i, \mathcal{P}^\star_i}} q^\prime$, then $\text{NN}$ satisfies the liveness property, i.e., $\text{NN}, \mathcal{X}_\text{init} \models \phi_{\text{liveness}}$. 
\end{thm}

In words, Theorem~\ref{thm:global_safe} guarantees that the global NN composed from provably safe local networks is still safe (i.e., satisfies $\phi_{\text{safety}}$). This reflects the fact that the composition of the global network respects the linear regions on which the local networks are defined. Moreover, if the local networks satisfy the \emph{local} reachability requirements, then the global NN satisfies the liveness property $\phi_{\text{liveness}}$, which reflects the fact that the set $\mathcal{Q}_{q, \mathcal{P}}$ in~\eqref{eq:Qp} is defined to guarantee progress towards the goal.

In practice, by combining the local networks into a single NN controller, it allows one to repair the NN controller in a systematic way when it fails to meet the liveness property. For example, if a local network $\text{NN}_q$ fails to satisfy the local reachability requirement $\mathrm{Reach}(q, \text{NN}_q) \subseteq \bigcup_{q^\prime \in \mathcal{Q}_{q, \mathcal{P}^\star}} q^\prime$, then only $\text{NN}_q$ need to be improved, such as by further training with augmented data collected at the state $q$, without affecting local networks that satisfy the specification at other abstract states.

\section{Experimental Results}  
\label{sec:results}
We evaluated the proposed framework both in simulation and on an actual robotic vehicle. All experiments were executed on an Intel Core i9 2.4-GHz processor with 32 GB of memory. 

\subsection{Controller Performance Comparison in Simulation}
We first present simulation results of a wheeled robot under the control of NN controllers trained by our algorithm. Let the state vector of the system be $x = [\zeta_x, \zeta_y, \theta]^\top \in \mathcal{X} \subset \mathbb{R}^3$, where $\zeta_x$, $\zeta_y$ denote the coordinates of the robot, and $\theta$ is the heading direction. The discrete-time dynamics of the robot is given by:
\begin{align}
    \zeta_x^{(t+\Delta t)} &= \zeta_x^{(t)} + \Delta t\ v\ \text{cos}(\theta^{(t)}) \notag \\
    \zeta_y^{(t+\Delta t)} &= \zeta_y^{(t)} + \Delta t\ v\ \text{sin}(\theta^{(t)}) \label{eq:dubin_car}\\
    \theta^{(t+\Delta t)} &= \theta^{(t)} + \Delta t\ u^{(t)} \notag
\end{align}
where the control input $u^{(t)} \in \mathbb{R}$ is determined by a neural network controller, i.e., $u^{(t)} = \text{NN}(x^{(t)})$, $\text{NN} \in \mathcal{P}^{K \times b} \subset \mathbb{R}^{1 \times 4}$ with the
controller space $\mathcal{P}^{K \times b}$ considered to be a hyperrectangle. We choose discrete time step size $\Delta t = 0.1$.

\begin{figure}[!t]
    \center
    \resizebox{.49\textwidth}{!}{
    \begin{tabular}{c|c}
        \rotatebox{90}{$\qquad\qquad\quad$\textbf{Workspace 1}} &
        \includegraphics[height=0.4\textwidth]{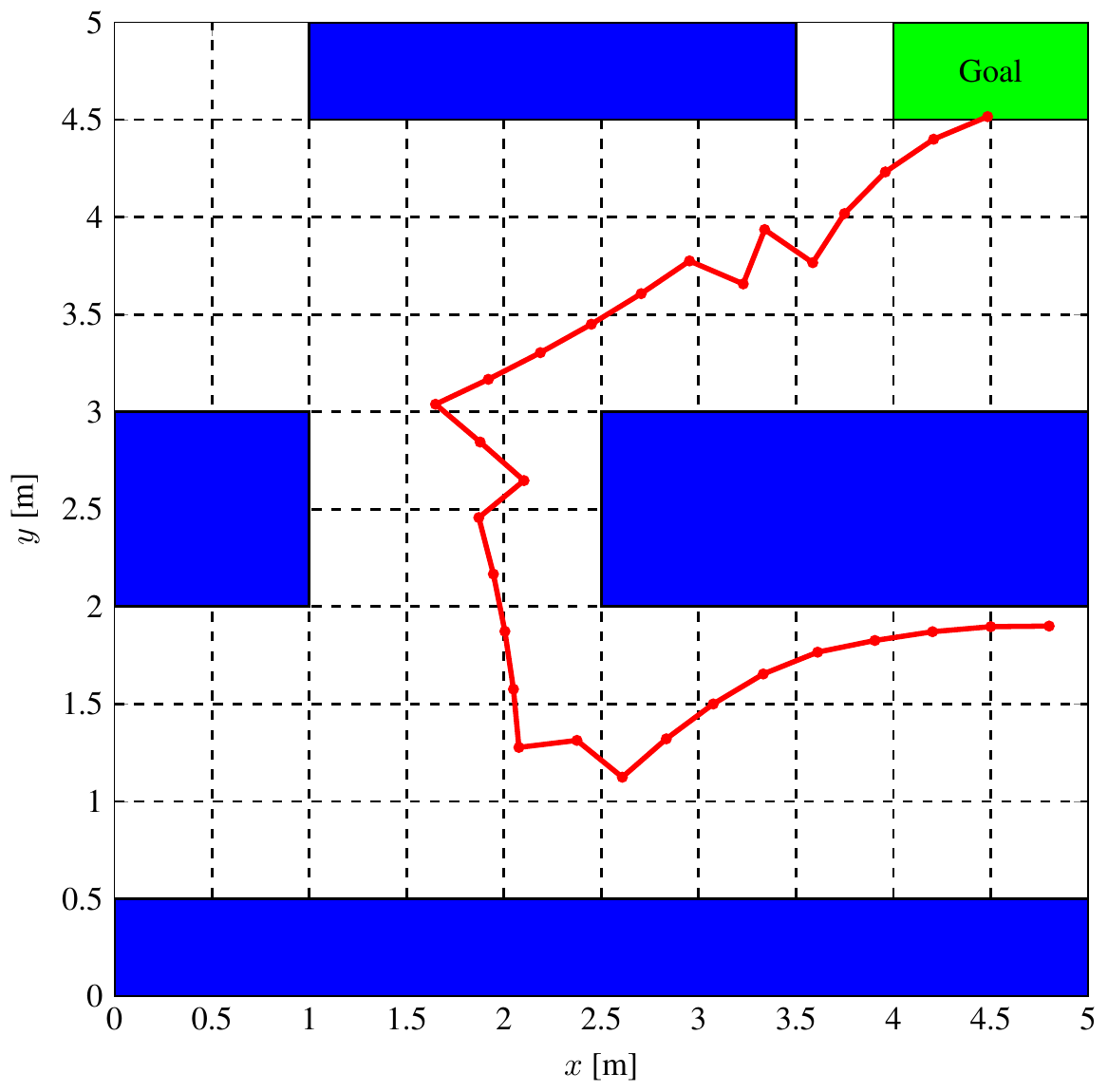}
        \includegraphics[height=0.4\textwidth]{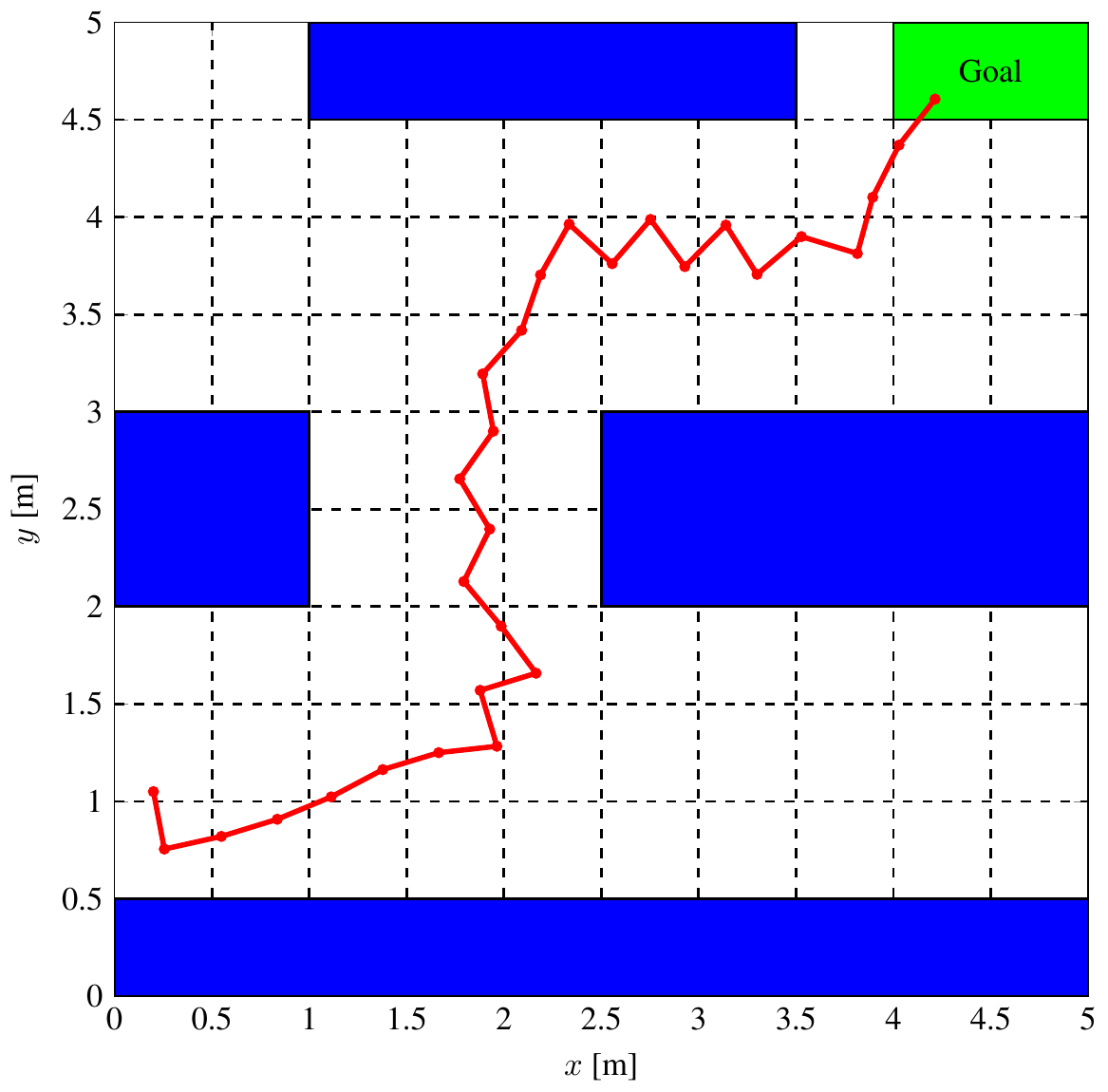}
        \includegraphics[height=0.4\textwidth]{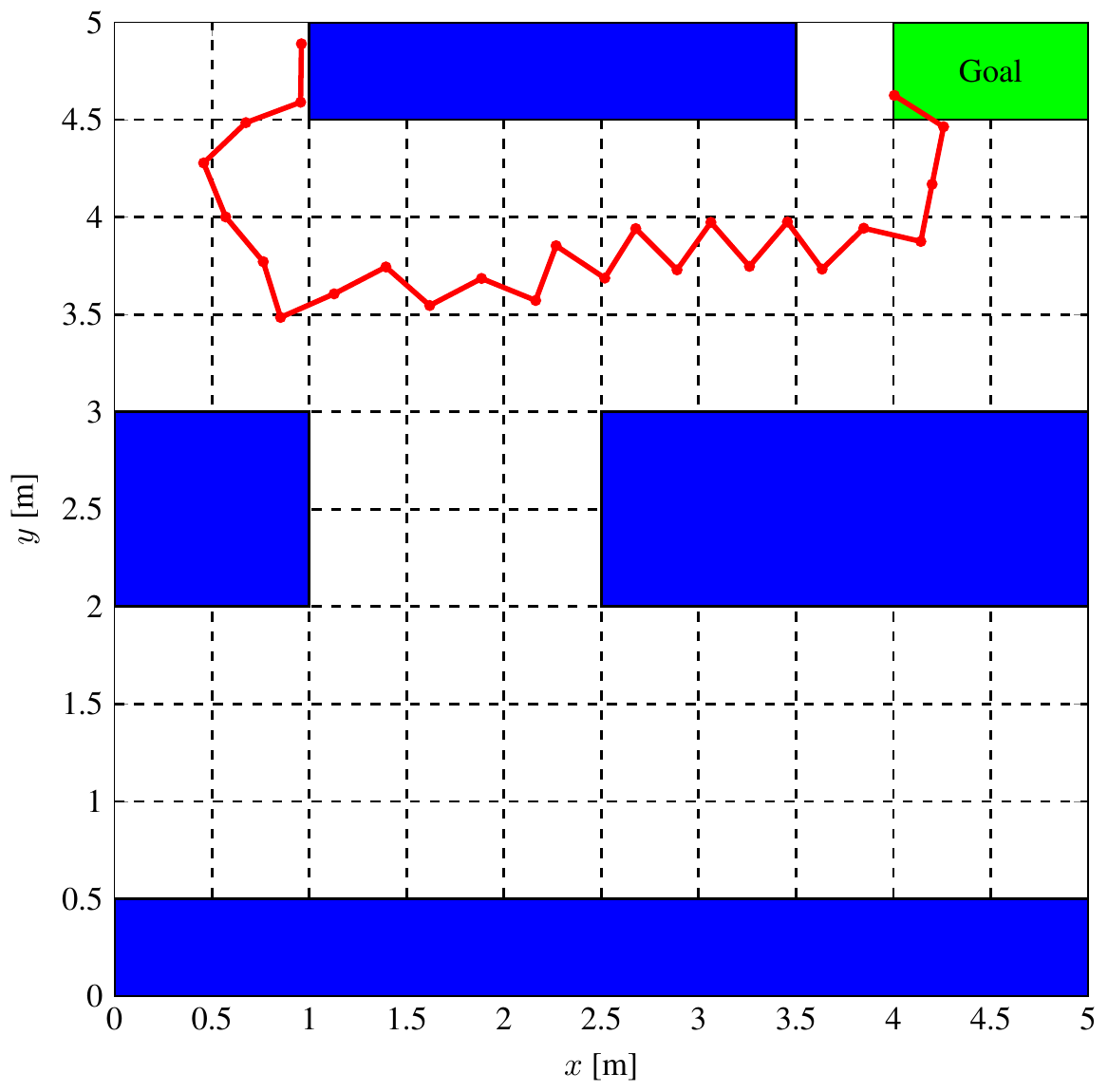} 
    \\ \hline
        \rotatebox{90}{$\qquad\quad\quad$\textbf{Workspace 2}} &
        \includegraphics[height=0.4\textwidth]{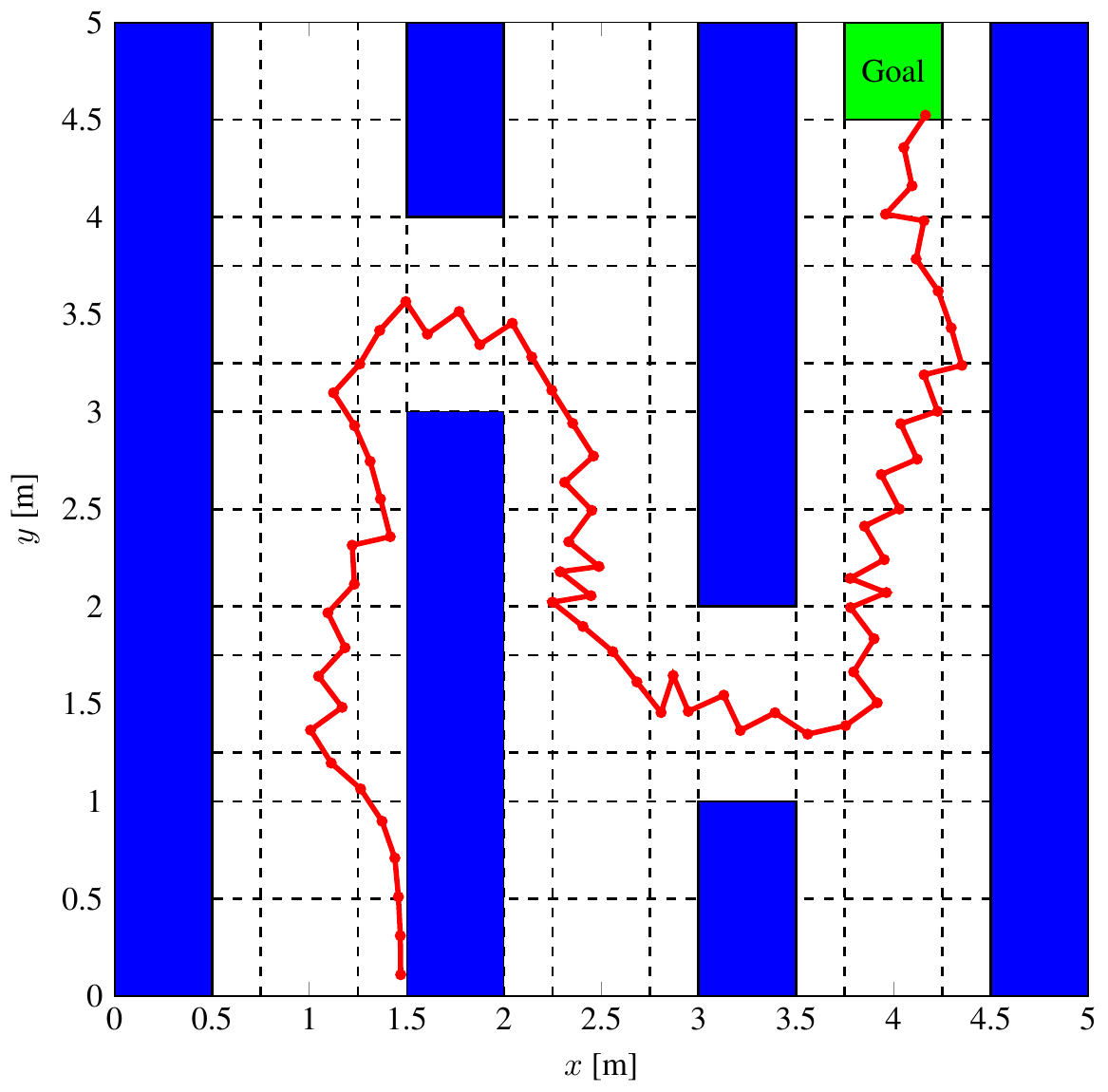}
        \includegraphics[height=0.4\textwidth]{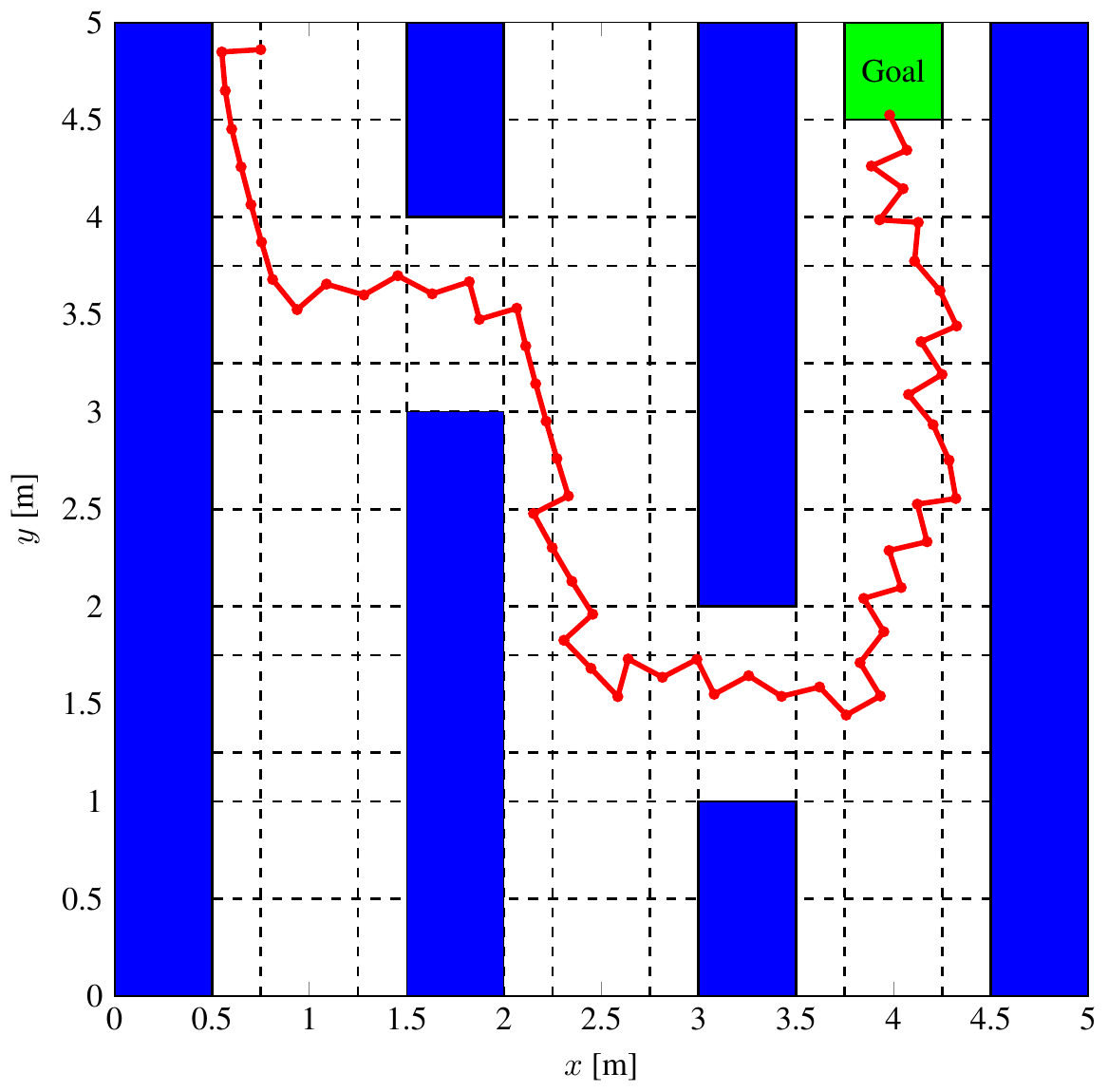}
        \includegraphics[height=0.4\textwidth]{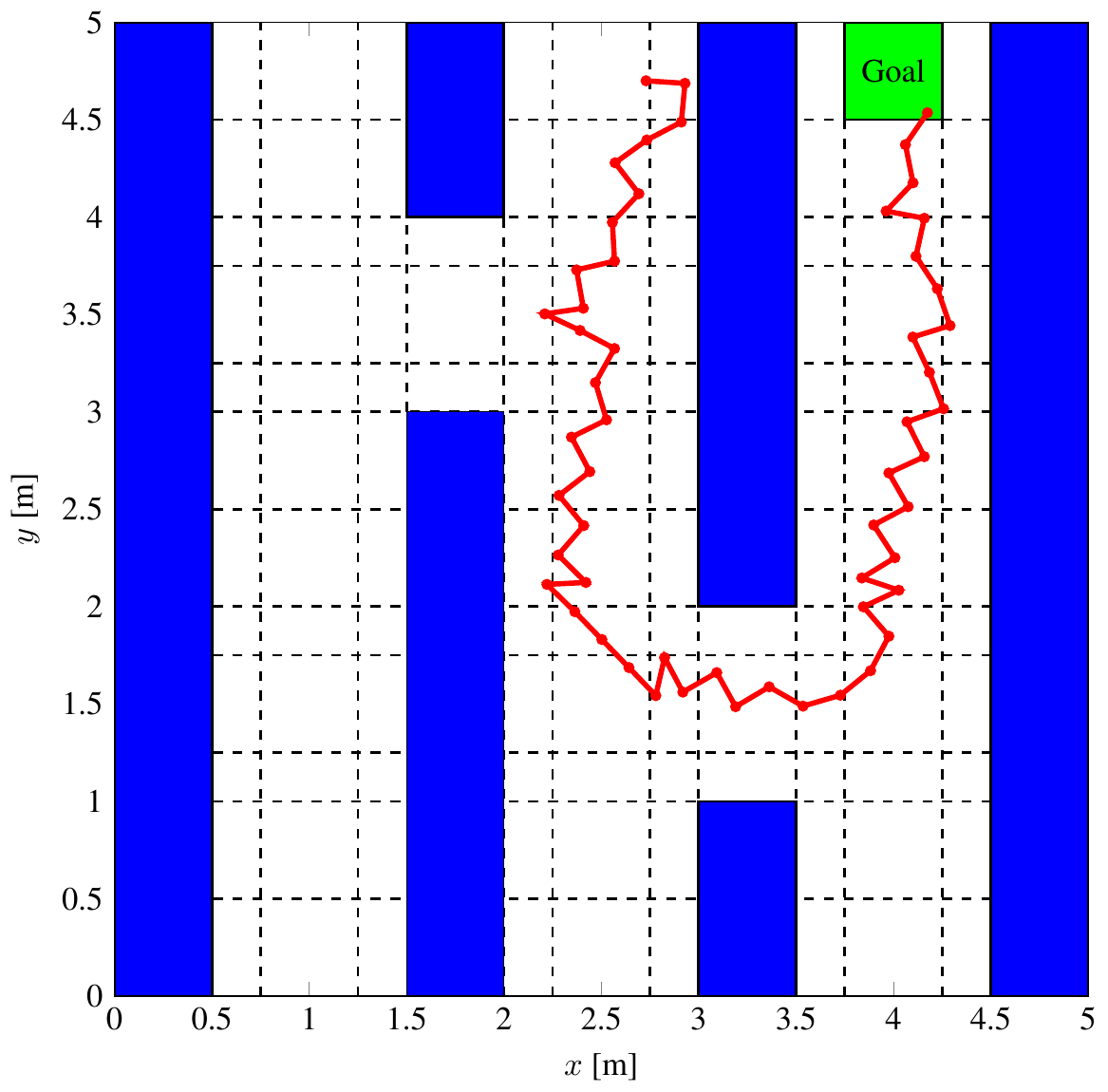}
    \end{tabular}
    }
    \caption{Workspace $1$ (the upper row) and workspace $2$ (the lower row) are partitioned into abstract states (dash lines) either uniformly or non-uniformly. Trajectories starting from different initial states satisfy both the safety specification $\phi_\text{safety}$ (blue areas are obstacles) and the liveness specification $\phi_\text{liveness}$ for reaching the goal (green area).}
    \label{fig:traj}  
\end{figure}

We considered two different workspaces indexed by $1$ and $2$ as shown in the upper and lower row of Figure~\ref{fig:traj}, respectively. As the first step of our algorithm, we discretized the state space $\mathcal{X} \subset \mathbb{R}^3$ and the controller space $\mathcal{P}^{K \times b} \subset \mathbb{R}^{1 \times 4}$ as described in Section~\ref{subsec:abst_model}. To illustrate the flexibility in the choice of partition strategies, we partitioned the state space corresponding to workspace $1$ uniformly into $552$ abstract states, while partitioning the state space corresponding to workspace $2$ non-uniformly into $904$ abstract states. In both cases, the range of heading direction $\theta \in [0, 2\pi)$ is uniformly partitioned into $8$ intervals, and the partitions of the $x$, $y$ dimensions are shown as the dashed lines in the workspaces in Figure~\ref{fig:traj}. We uniformly partitioned $\mathcal{P}^{K \times b}$ into $160$ hyperrectangles. By computing the reachable sets using the reachability tool TIRA~\citep{tira}, we constructed the posterior graph $S_\mathrm{Post}$, which is then used to find the set of safe abstract states $\mathbb{X}_\text{safe}$ and the function $P_\text{safe}$. The execution time to compute the posterior graph and to identify the set of safe states can be found in Table~\ref{tab:partition}.



We used the nonlinear MPC solver FORCES~\citep{forcespro,forcesnlp} as an expert to collect training data, and used Keras~\citep{chollet2015keras} to train a shallow NN (one hidden layer) with $4$ hidden layer neurons for each abstract state $q \in \mathbb{X}_\text{safe}$. At the end of training, we projected the trained NN weights only once as mentioned in Section~\ref{subsec:project_weights}. For workspace $1$, it takes $367$ seconds to collect all the training data, and $463$ seconds to train all the local NNs including the projection of the NN weights. For workspace $2$, the execution time for collecting data is $601$ seconds, and the total time for training and projection is $695$ seconds. 

In Figure~\ref{fig:traj}, we show some trajectories under NN controllers trained by our algorithm in both workspaces. Despite we choose trajectories with initial states in the set $\mathcal{X}_\text{init} = \bigcup_{q \in \mathbb{X}_\text{safe}} q$ to be close to the obstacles or initially heading towards the obstacles, all the trajectories are collision-free as guaranteed by our algorithm. Moreover, by assigning controller partitions based on strategies in Section~\ref{subsec:assign_partition}, all trajectories satisfy the liveness specification $\phi_\text{liveness}$.  

\begin{figure}[!t]
    \center
    \resizebox{.49\textwidth}{!}{
    \begin{tabular}{c|c}
        \rotatebox{90}{$\quad$\textbf{Standard Imitation Learning}}
         & 
        \includegraphics[height=0.4\textwidth]{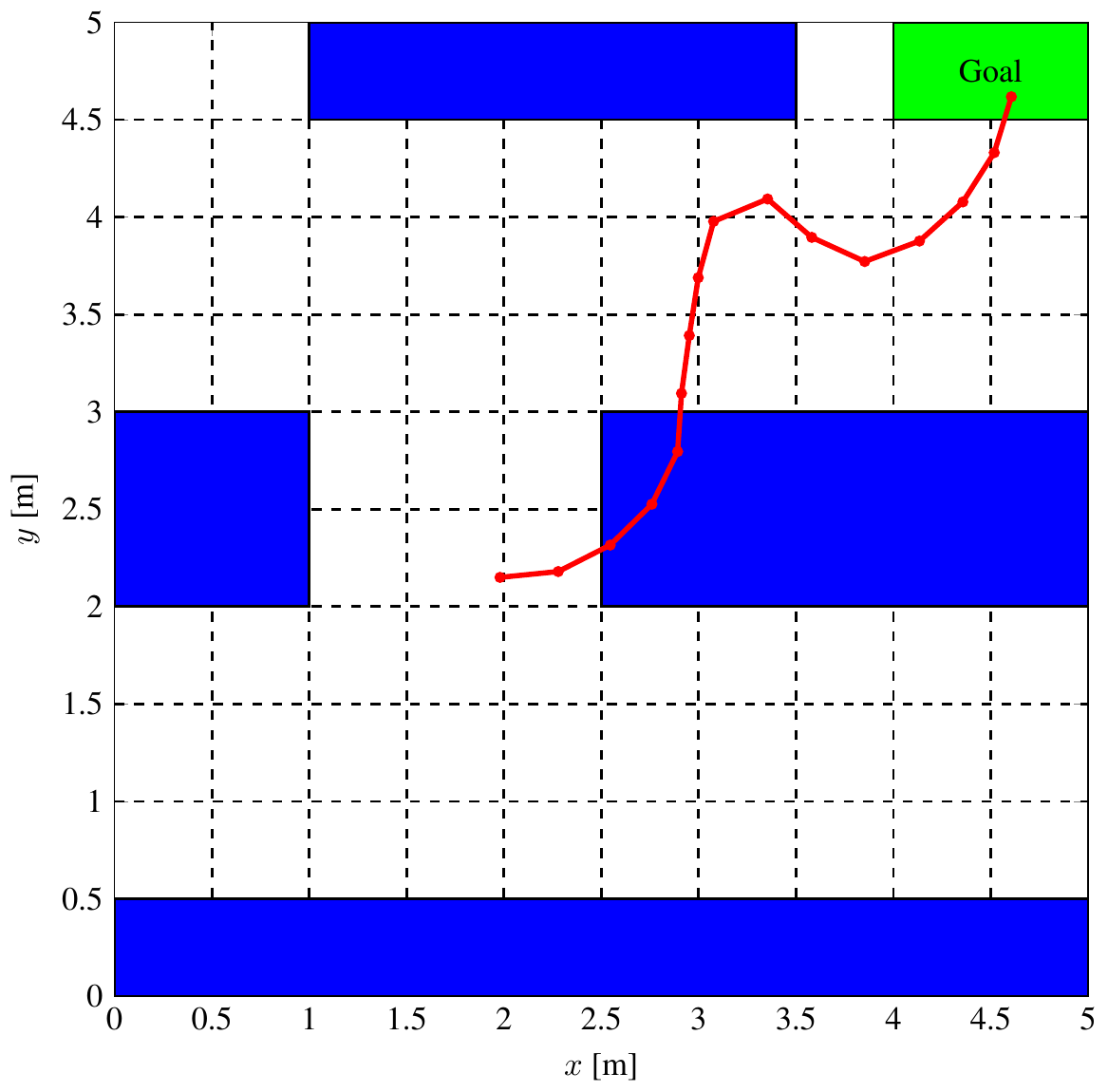}
        \includegraphics[height=0.4\textwidth]{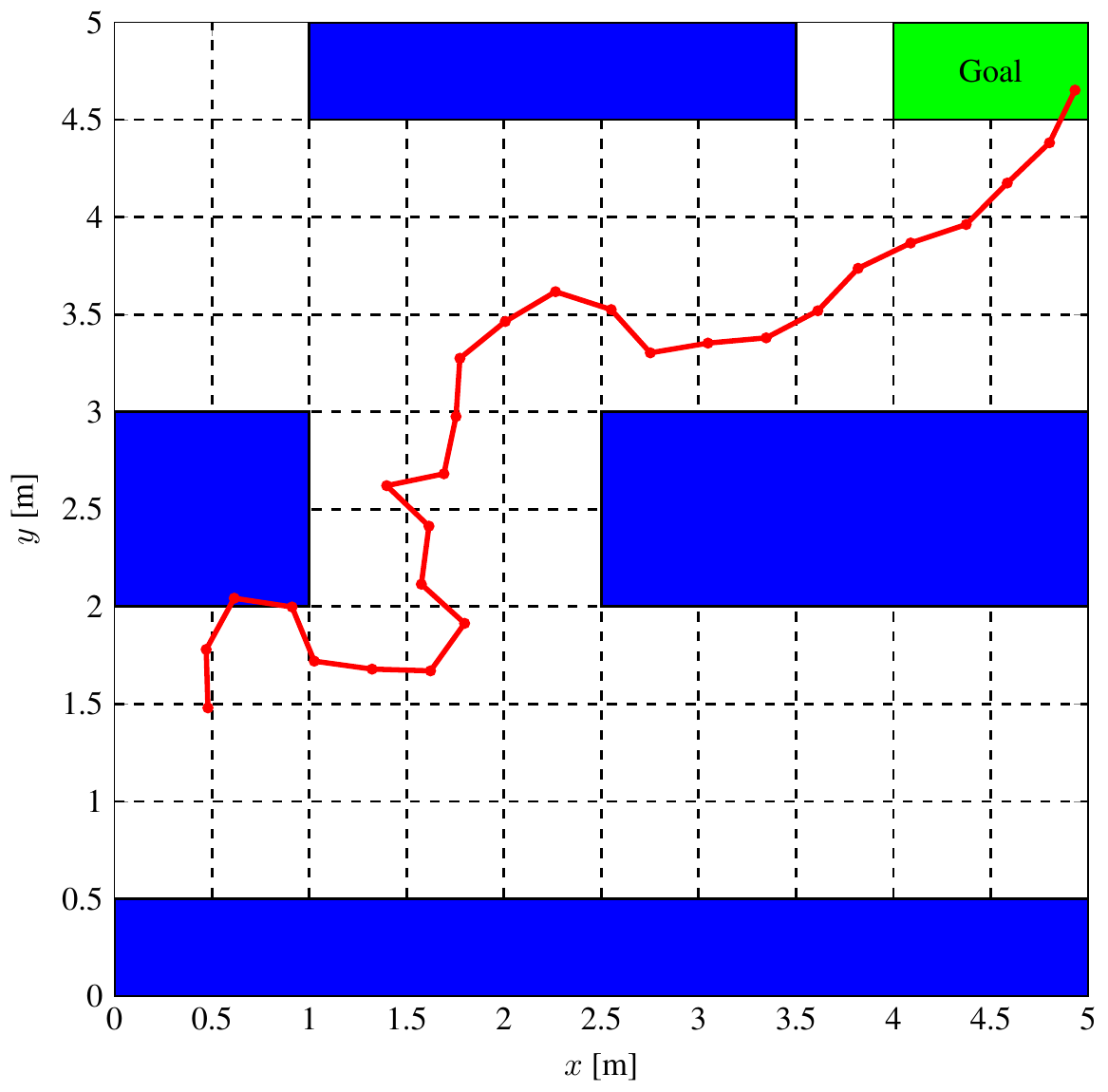} 
        \includegraphics[height=0.4\textwidth]{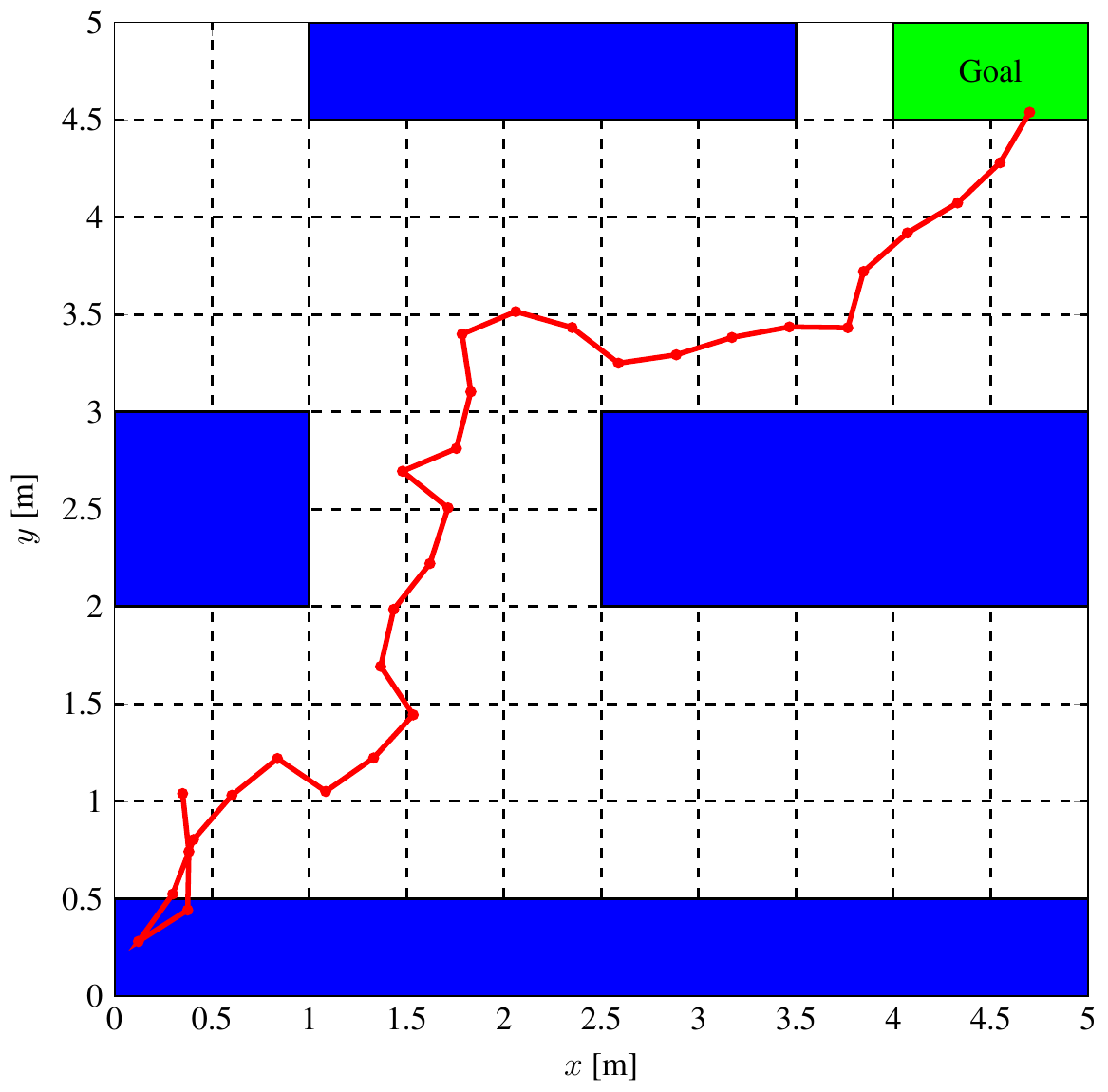}
        \\ \hline
        \rotatebox{90}{$\qquad$\textbf{Provably Safe Training}}
         & 
        \includegraphics[height=0.4\textwidth]{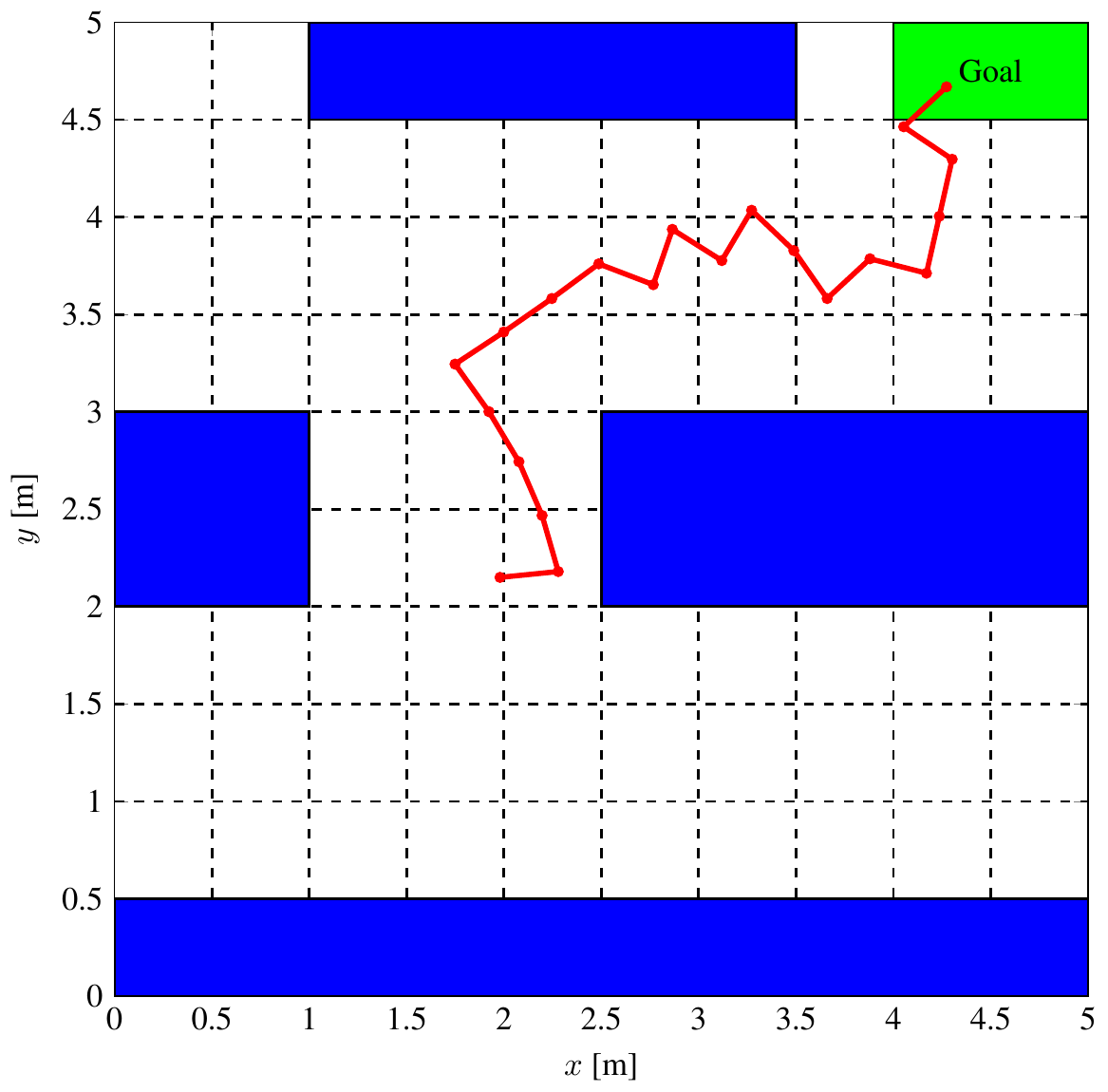}
        \includegraphics[height=0.4\textwidth]{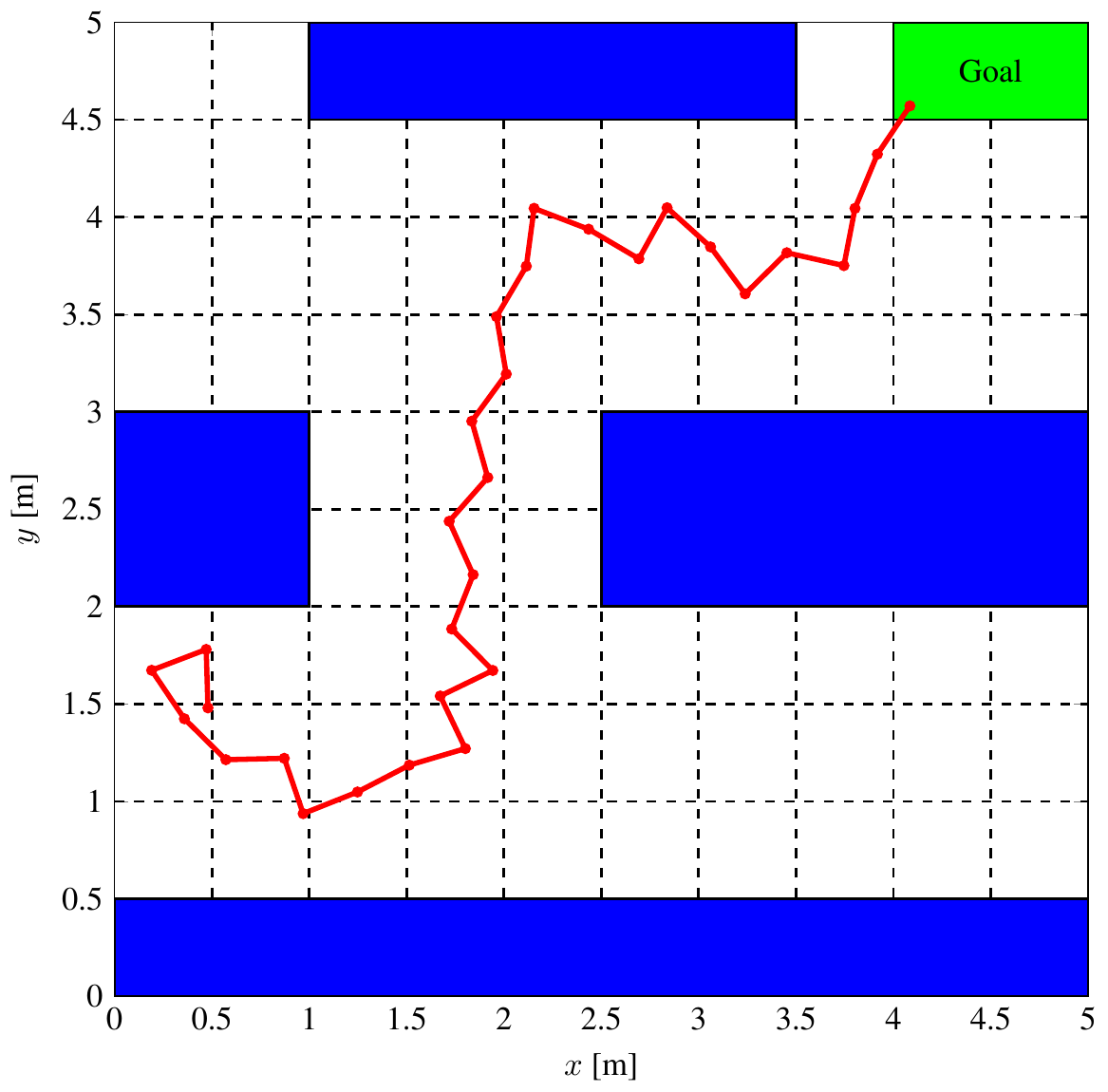}
        \includegraphics[height=0.4\textwidth]{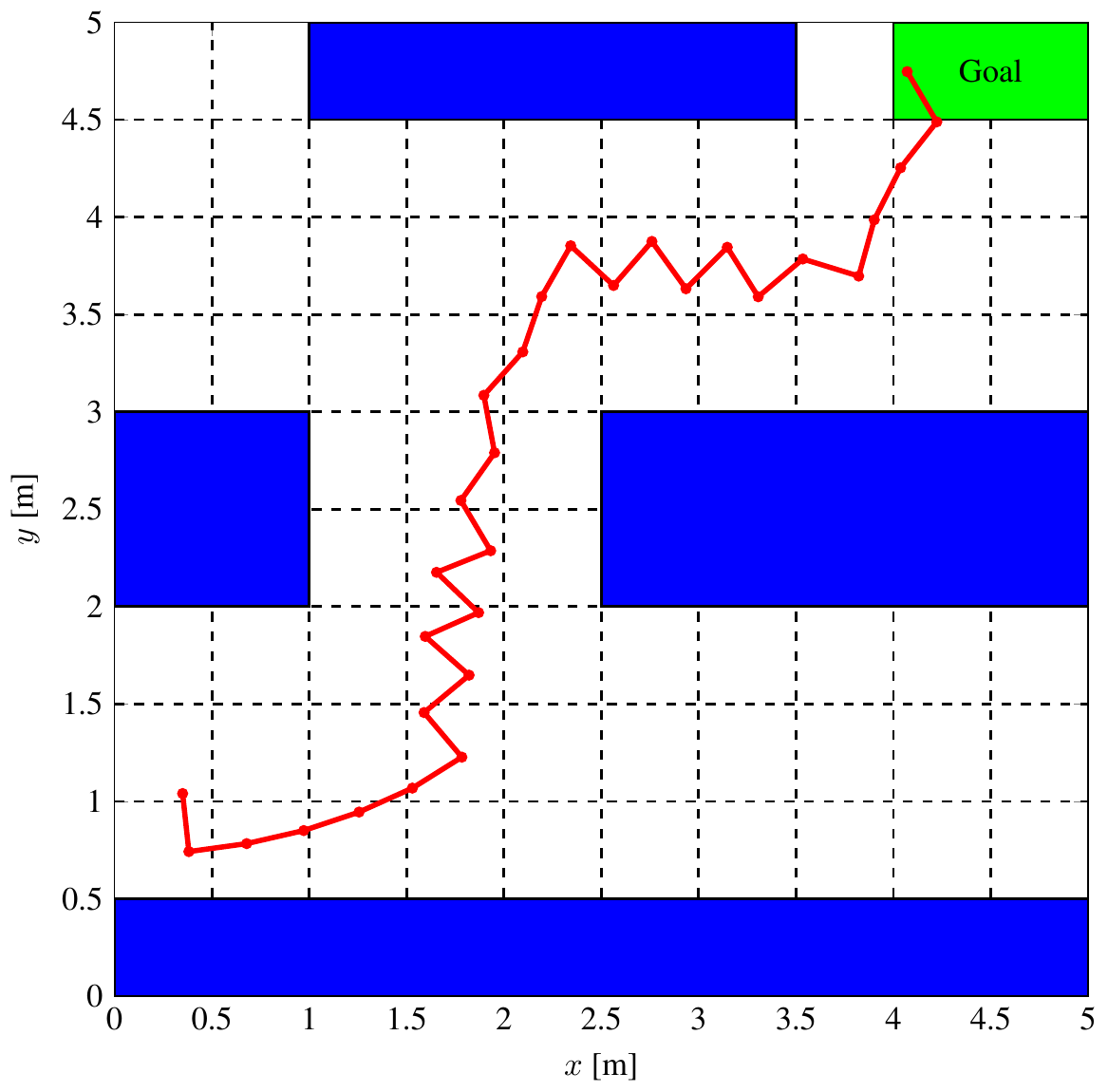}
    \end{tabular}  
    }  
    \caption{The upper row shows trajectories resulting from NN controllers trained using standard imitation learning, where the NN architectures are (left) $2$ hidden layers with $10$ neurons per layer, (middle) $2$ hidden layers with $64$ neurons per layer, and (right) $3$ hidden layers with $128$ neurons per layer. The lower row shows trajectories resulting from NN controllers trained using our algorithm. With the same initial states (two sub-figures in the same column), only the NN controllers trained by our algorithm result into collision-free trajectories.} 
    \label{fig:compare}  
\end{figure}     

Next, we compare NN controllers trained by our algorithm with those trained by standard imitation learning, which minimizes the regression loss without taking into account the safety guarantee. All NN controllers are trained using the same set of the training data. We vary the NN architectures for the NNs trained by standard imitation learning to achieve better performance, and train them using enough episodes for the loss to be low enough. Nevertheless, as shown in Figure~\ref{fig:compare}, for all the NN controllers trained by standard imitation learning, we are able to find initial states starting from which the trajectories are not safe. However, with the same initial states, trajectories under NN controllers trained by our algorithm are collision-free as the framework guarantees. 

\subsection{Actual Robotic Vehicle}  
\begin{figure}[!ht]  
    \center
    \resizebox{.49\textwidth}{!}{
        \subfloat[]{\raisebox{3ex}{\includegraphics[width=0.38\linewidth]{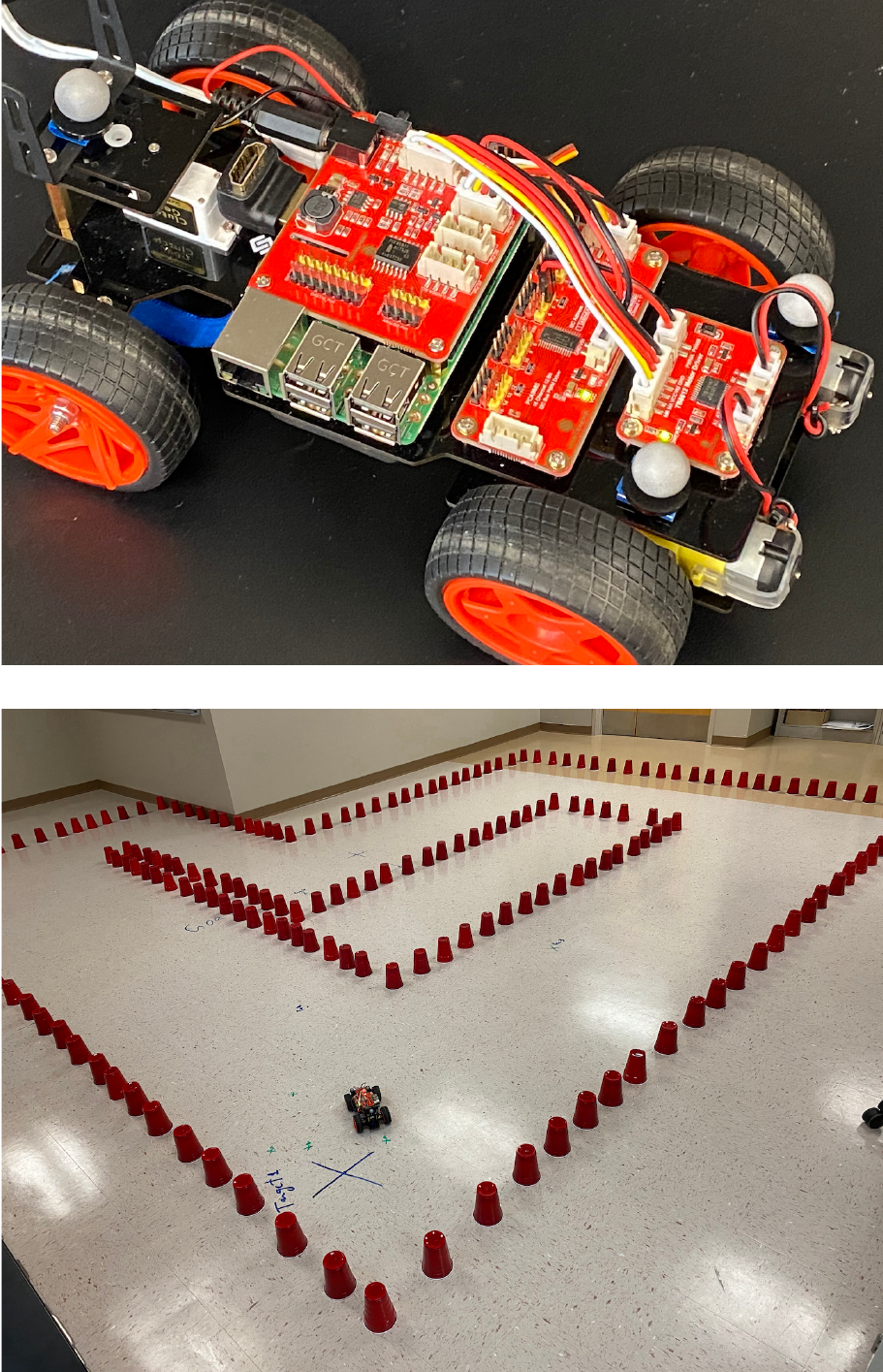}}}
        \subfloat[]{\includegraphics[width=0.6\linewidth]{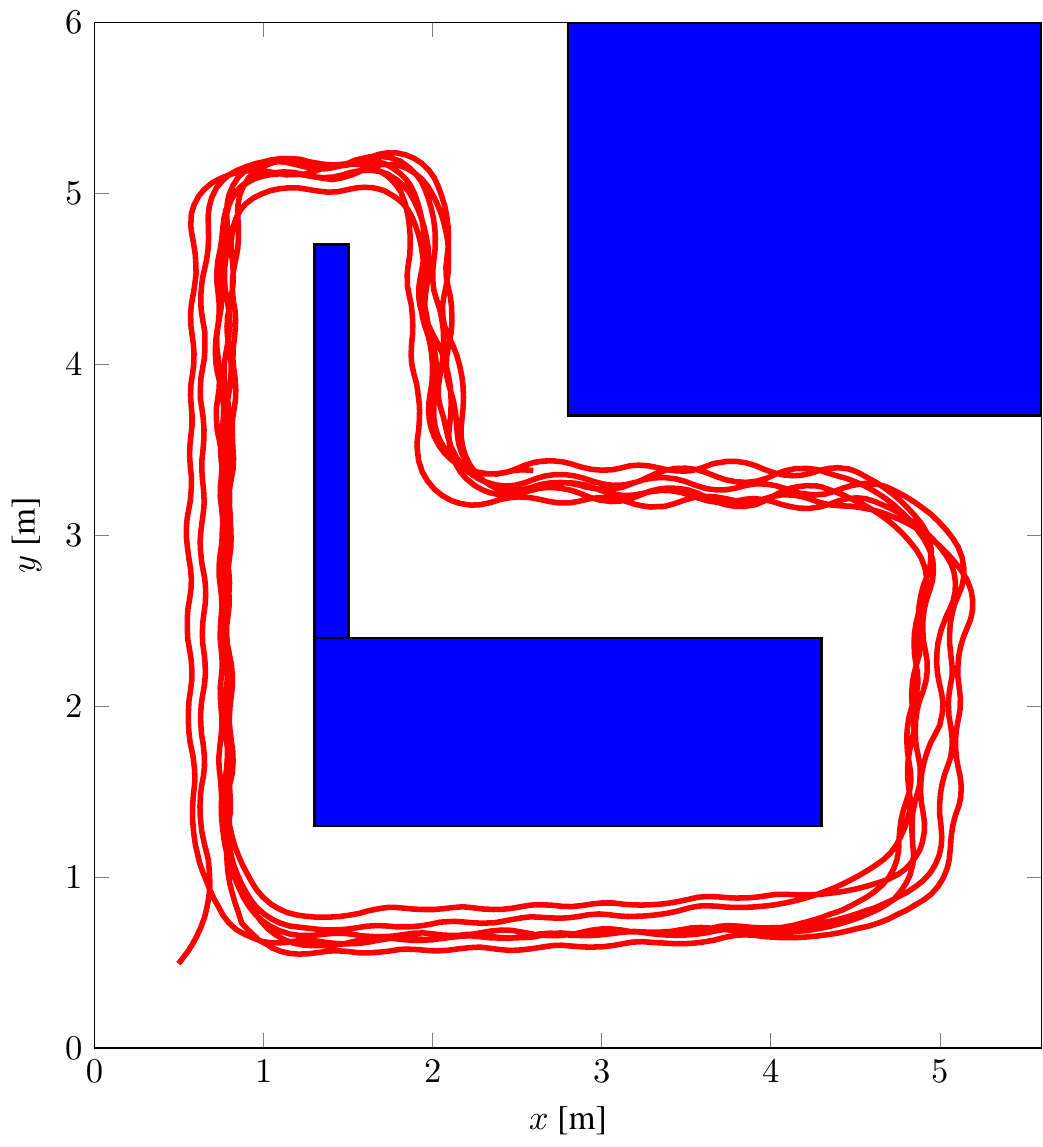}}  
    }
    \caption{(a) PiCar and workspace (b) The PiCar's trajectory (red) on the race track satisfies the safety property $\phi_\text{safety}$. } 
    \label{fig:picar} 
\end{figure}

We tested the proposed framework on a small robotic vehicle called PiCar, which carries a Raspberry Pi that runs the NN controller trained by our algorithm. We used the motion capture system Vicon to measure states of the PiCar in real time. Figure~\ref{fig:picar} (a) shows the PiCar and our experimental setup. We modeled the PiCar dynamics by the rear-wheel bicycle drive~\citep{rearwheel}, and took model-error into consideration by adding a bounded disturbance when computing the reachable sets. We trained the local networks $\text{NN}_q$ using  Proximal Policy Optimization (PPO)~\citep{ppo}. Figure~\ref{fig:picar} (b) shows the PiCar's trajectory under the NN controller trained by our framework. The PiCar runs on the race track for multiple loops and satisfies the safety property $\phi_\text{safety}$. 

\subsection{Scalability Study} 
\noindent\textbf{1- Scalability with respect to Partition Granularity:} 
We first show scalability of our algorithm with respect to the choice of partition parameters. Using the two workspaces in Figure~\ref{fig:traj}, we increase the number of abstract states and controller partitions by decreasing the partition grid size and report the execution time for each part of our framework in Table~\ref{tab:partition}. We notice that the execution time grows linearly with the number of abstract states and the number of controller partitions.

Although we conducted all the experiments on a single CPU core, we note that our algorithm can be highly parallelized. For example, computing reachable sets of the abstract states, checking intersection between the posteriors and the abstract states when constructing the posterior graph, and training local neural networks $\text{NN}_q$ can all be done in parallel. After training the NN controller, the execution time of the controller is almost instantaneous, which is a major advantage of NN controllers.

\begin{table}[!ht]
    \caption{Scalability with respect to Partition Granularity} 
    \begin{center}
    \resizebox{.99\columnwidth}{!}{
    \begin{tabular}{|c|c|c|c|c|c|c|c|}
    \hline
    \textbf{Workspace} & \textbf{Number of} & \textbf{Number of}  & \textbf{Number of}         & \textbf{Compute}   &\textbf{Construct} &\textbf{Compute}             &\textbf{Assign} \\ 
    \textbf{Index}     & \textbf{Abstract}  & \textbf{Controller} & \textbf{Safe \& Reachable} & \textbf{Reachable} &\textbf{Posterior} &\textbf{Function}            &\textbf{Controller} \\
    \textbf{}          & \textbf{States}    & \textbf{Partitions} & \textbf{Abstract States}   & \textbf{Sets [s]}  &\textbf{Graph [s]} &\textbf{$P_\text{safe}$ [s]} &\textbf{Partitions [s]} \\
    \hline\hline
    1 & 552  & 160 & 400  & 52.6  & 82.3   & 0.06 & 0.7 \\ \cline{1-8} 
    1 & 552  & 320 & 400  & 107.5 & 160.3  & 0.1  & 0.9 \\ \cline{1-8}
    1 & 552  & 640 & 400  & 223.1 & 329.6  & 0.2  & 1.7 \\ \cline{1-8}
    1 & 1104 & 160 & 800  & 108.2 & 333.0  & 0.2  & 2.3 \\ \cline{1-8}
    1 & 1104 & 320 & 800  & 219.6 & 684.2  & 0.4  & 2.7 \\ \cline{1-8}
    1 & 1104 & 640 & 800  & 451.5 & 1297.4 & 0.6  & 4.2 \\ \cline{1-8}
    2 & 904  & 160 & 632  & 88.1  & 159.1  & 0.1  & 1.0 \\ \cline{1-8} 
    2 & 904  & 320 & 632  & 203.6 & 313.2  & 0.2  & 1.1 \\ \cline{1-8} 
    2 & 904  & 640 & 632  & 393.2 & 660.8  & 0.3  & 1.7 \\ \cline{1-8}  
    2 & 1808 & 160 & 1264 & 202.1 & 634.6  & 0.3  & 3.4 \\ \cline{1-8} 
    2 & 1808 & 320 & 1264 & 388.6 & 1298.1 & 0.6  & 4.0 \\ \cline{1-8} 
    2 & 1808 & 640 & 1264 & 778.2 & 2564.4 & 0.9  & 5.9 \\ \cline{1-8} 
\end{tabular}       
}
\end{center}
\label{tab:partition} 
\end{table}

\noindent\textbf{2- Scalability with respect to System Dimension:} 
Abstraction-based controller design is known to be computational expensive for high-dimensional systems due to the curse of dimensionality. In Table~\ref{tab:linear}, we show scalability of our algorithm with respect to the system dimension. To conveniently increase the system dimension, we consider a chain of integrators represented as the linear system $x^{(t+1)} = Ax^{(t)} + Bu^{(t)}$, where $A \in \mathbb{R}^{n \times n}$ is the identity matrix, and $u^{(t)} \in \mathbb{R}^2$. With the fixed number of controller partitions, 
Table~\ref{tab:linear} shows that the number of abstract states and execution time grow exponentially with the system dimension $n$. Nevertheless, our algorithm can handle a high-dimensional system in a reasonable amount of time. 

\begin{table}[!ht]
    \caption{Scalability with respect to System Dimension} 
    \begin{center} 
    \resizebox{.99\columnwidth}{!}{
    \begin{tabular}{|c|c|c|c|c|c|c|c|}
    \hline
    \textbf{System} & \textbf{Number of} & \textbf{Compute Reachable} &\textbf{Construct Posterior} \\ 
    \textbf{Dimension $n$} & \textbf{Abstract States} & \textbf{Sets [s]} &\textbf{Graph [s]} \\ 
    \hline\hline
    2  & 69    & 0.6   & 0.7 \\ \cline{1-4}
    4  & 276   & 2.7   & 2.6 \\ \cline{1-4} 
    6  & 1104  & 11.7  & 34.2 \\ \cline{1-4}
    8  & 4416  & 57.1  & 521.0 \\ \cline{1-4}
    10 & 17664 & 258.1 & 9840.4 \\ \cline{1-4}
\end{tabular}       
}
\end{center}
\label{tab:linear} 
\end{table}

\section{Conclusion} 
In this paper, we proposed a framework for training neural network controllers for nonlinear dynamical systems with rigorous safety and liveness guarantees. Our framework computes a finite-state abstract model that captures the behavior of all possible CPWA controllers, and enforces the correct behavior during training through a NN weight projection operator, which can be applied on top of any existing training algorithm. We implemented our framework on an actual robotic vehicle that shows the capability of our approach in real-world applications. Compared to existing techniques, our framework results in provably correct neural network controllers removing the need for online monitoring, predictive filters, barrier functions, or post-training formal verification.
Future research directions include extending the framework to multiple agents in unknown environment, and using the proposed abstract model to generalize learning for different tasks in meta-learning setting.  



\bibliographystyle{apalike}
\bibliography{biblio}  



\end{document}